\documentclass[pra,twocolumn,superscriptaddress]{revtex4}

\usepackage{mathbbol}
\usepackage{amsmath,amssymb}
\usepackage{amssymb,enumerate}
\usepackage{graphicx}
\usepackage{graphics}
\usepackage{amsmath}
\usepackage{amsthm,bbm}
\usepackage{color}
\usepackage{dsfont}
\usepackage{hyperref}
\usepackage{bbm}

\usepackage{lipsum}
\usepackage{lmodern}
\usepackage{tcolorbox}

\usepackage{amsfonts}
\usepackage{graphicx,graphics,epsfig,times,bm,bbm,amssymb,amsmath,amsfonts,mathrsfs}
\usepackage[normalem]{ulem}
\usepackage{setspace}
\usepackage{subfigure}
\usepackage{dsfont}
\usepackage{braket}
\usepackage{upgreek }
\usepackage{tikz}
\usepackage{natbib}
\usepackage{chngcntr}

\newtheorem{theorem}{Theorem}

\newtheorem{corollary}[theorem]{Corollary}

\newtheorem{lemma}{Lemma}

\newtheorem{proposition}{Proposition}

\newcommand{\bes} {\begin{subequations}}
\newcommand{\ees} {\end{subequations}}
\newcommand{\bea} {\begin{eqnarray}}
\newcommand{\eea} {\end{eqnarray}}
\newcommand{\be} {\begin{equation}}
\newcommand{\ee} {\end{equation}}

\def\i{{\textbf{x}}}
\def\m{\text{max}}

\def\>{\rangle}
\def\<{\langle}
\def\Tr{\textrm{Tr}}

\newcommand{\ignore}[1]{}

\usepackage{amsmath,amssymb}

\begin{document}	
	\title{Inaccessible information in  probabilistic models of quantum systems, non-contextuality inequalities and noise thresholds for contextuality}

	\author{Iman Marvian}
\affiliation{Departments of Physics \& Electrical and Computer Engineering, Duke University, Durham, North Carolina 27708, USA}
	\begin{abstract}
Classical probabilistic models of (noisy) quantum systems are not only relevant for understanding the non-classical features of quantum mechanics, but  they are also useful for determining  the possible advantage of using quantum resources for information processing tasks.  A common feature of these models is the presence of inaccessible information, as captured by the concept of preparation contextuality:   There are  ensembles of quantum states  described by the same density operator, and hence operationally indistinguishable, and yet in any probabilistic (ontological) model, they should be described by distinct  probability distributions.  In this work, we  quantify the inaccessible information of a model in terms of the maximum distinguishability of probability distributions associated to any pair of  ensembles with identical density operators, as quantified by the total variation distance of the distributions.  We obtain a family of lower bounds on this maximum distinguishability in terms of experimentally measurable quantities.  In the case of an ideal qubit this leads to a lower bound of, approximately, $0.07$. These bounds can also  be interpreted as a new class of robust preparation non-contextuality inequalities.  Our non-contextuality inequalities are phrased in terms of generalizations of max-relative entropy and trace distance for general operational theories, which  could be of independent interest.

Under sufficiently strong noise any quantum system becomes preparation non-contextual, i.e., can be described by models with zero inaccessible information. Using our non-contextuality inequalities,  we show that this can happen only if the noise channel has the average gate fidelity less than or equal to $D^{-1}(1+2^{-1}\cdots+D^{-1})$,  where $D$ is the dimension of the Hilbert space. 

	\end{abstract}
	\maketitle	

\section{Introduction}

Bell's groundbreaking work \cite{bell1964einstein} in 1964  not only clarified  astonishing features of quantum entanglement, it also introduced a paradigm for probing and understanding  properties of nature,  independent of  the formalism of quantum mechanics. In this paradigm one assumes there is a classical probabilistic model, also known as an \emph{ontological} model,  which describes 
the experiment under consideration and  satisfies certain physically-motivated properties \cite{harrigan2010einstein}. In such models each quantum state  $\psi$ corresponds to a probability distribution over the possible values of some hidden variables, also known as  the  \emph{ontic} states. The ontic state $\lambda$ determines the outcome of any measurement in a deterministic or stochastic fashion. Then,  assuming the model satisfies the  desirable properties,  such as locality in the case of Bell's inequalities, one obtains non-trivial  constraints   on the possible observable statistics.  Observing violation of these constraints in an actual experiment  reveals properties of nature which remain valid and meaningful,  independent of the validity of quantum mechanics. 

 Beside its foundational significance, this paradigm turns out to be   useful for understanding the power of quantum mechanical systems for information processing tasks. For instance, in an interesting twist, it was found that violation of Bell's inequalities can be used for device-independent quantum key distribution, where it is possible to achieve information-theoretic security without trusting the used quantum devices \cite{mayers1998quantum, barrett2005no, acin2006bell, acin2007device}.

In the case of Bell's inequalities, the imposed constraint on the model  is  a certain notion of locality  in bipartite systems, namely lack of  superluminal causal influences.  One can consider other physical properties which should be satisfied by any  reasonable physical theory. One such property is non-contextuality, originally introduced by Bell \cite{bell1966problem} and Kochen and Specker \cite{kochen1975problem}, which was later generalized by Spekkens \cite{spekkens2005contextuality}. Roughly speaking, the principle of generalized non-contextuality, which is sometimes motivated by the Leibniz's principle of identity of indiscernibles \cite{spekkens2019ontological}, states that  any two operationally indistinguishable scenarios should  have the same descriptions in the model (See Sec.\ref{Sec:Non}). An ideal (noiseless) quantum mechanical system is contextual, i.e.,  any ontological model of the system violates  non-contextuality   \cite{spekkens2005contextuality}.

It has been argued that contextuality captures several notions of non-classicality such as  anomalous weak values \cite{pusey2014anomalous} and negativity of quasiprobability representations \cite{spekkens2008negativity}. Furthermore, the significance of contextuality as a resource for information processing tasks has been extensively studied, e.g. in the context of quantum computation \cite{howard2014contextuality, raussendorf2013contextuality, delfosse2015wigner, spekkens2007evidence}, cryptography \cite{spekkens2009preparation, banik2015limited, ambainis2019parity, spekkens2007evidence} and state discrimination  \cite{schmid2018contextual}.

\subsection*{Summary of Results}

In this work, we take an information-theoretic approach to the study of  contextuality. By definition \cite{spekkens2005contextuality},  a model is preparation contextual, if there are distinct ensembles of states which yield the same average density operator, and yet in the model they are represented by different probability distributions.  This means that by  preparing the system in one of these ensembles,  one can encode information in the ontic state $\lambda$ and this encoded information remains completely inaccessible by any physical measurements. We quantify this \emph{inaccessible information} of a model by considering the maximum distinguishability of pairs of  distributions associated to operationally indistinguishable ensembles, as quantified by the  total variation distance of the distributions (See definition in Eq. (\ref{kgkg}) and Eq. (\ref{kgkg2})).  
 This quantity, which is bounded between zero and one, quantifies deviation from preparation non-contextuality and has a simple interpretation: it determines the probability that a hypothetical observer who can directly observe the value of the ontic state $\lambda$ can distinguish two ensembles which are operationally indistinguishable.  
 
Our first main result is a set of lower  bounds on the inaccessible information in terms of experimentally measurable quantities (Theorems \ref{Thm000} and \ref{Thm00}). In the case of a single noiseless qubit, we show that for certain experimental setups the lower bound on the inaccessible information is, $(2-\sqrt{2})/{8} \approx 0.07$. 
 On the other hand, we find that for the  Kochen-Specker model of a qubit \cite{kochen1975problem}  this quantity is upper bounded by 0.5 (See Sec.\ref{Sec:Kochen}). Therefore, while the lowest possible value of the inaccessible information for a single noiseless qubit remains unknown, we find that its value is in the interval $[0.07, 0.5]$. 

The fact that the lower bound on the inaccessible information is non-zero for certain  experimental setups,  immediately gives a proof of contextuality of quantum mechanics. Furthermore,  setting the inaccessible information  equal to zero in these inequalities, we find a new class of robust non-contextuality inequalities. In general, robust non-contextuality inequalities, which are counterparts of Bell's inequalities,  impose non-trivial constraints  on the observable statistics based on the  assumption of preparation non-contextuality, and are violated in actual experiments \cite{mazurek2016experimental}, even in the presence of finite noise and imperfections  (See e.g. \cite{mazurek2016experimental,  schmid2018contextual, kunjwal2015kochen, schmid2018all, hameedi2017communication}). Our non-contextuality inequalities yield, as a special case,  a previously known  non-contextuality inequality \cite{mazurek2016experimental}.  Furthermore, we find a simple interpretation of these non-contextuality inequalities in terms of a family of guessing games (See Sec \ref{Sec:lkhf}). Our non-contextuality inequalities are phrased in terms of generalizations of  max-relative entropy \cite{datta2009min} and trace distance for general operational theories, which quantify  the distinguishability of  preparations, and  could be of independent interest  (See Sec.\ref{Sec:dist}).

Finally, we study contextuality in the presence of noise (See Sec.  \ref{Sec:noise}). It turns out that under sufficiently strong noise, quantum mechanical systems become non-contextual, i.e., can be described by non-contextual models. To study this phenomenon,  we assume the noise can be described by a quantum channel. Note that unlike the above results, which hold independently of the validity of quantum mechanics, here we study the problem in the framework of quantum mechanics.  

Our second main result, which is a corollary of our non-contextuality inequalities, is a noise threshold for contextuality: We show that for a system with Hilbert space of dimension $D$, if the noise is described by a quantum channel with the average gate fidelity larger than $D^{-1}(1+2^{-1}+\cdots+D^{-1})$, and assuming the noise channel is a one-to-one function, then it is still possible to perform prepare-measure experiments  demonstrating   preparation contextuality  (See Eq.(\ref{kjhjhkljhwilv})). In the case of a single qubit with depolarizing noise channel we show that this bound is tight. Furthermore, we find that there is a distinct (higher) noise level, above which the theory satisfies both preparation and measurement non-contextuality; namely, this happens when the noise channel is entanglement-breaking  (See Sec. \ref{Sec:noise}).

\section{Preliminaries}
The central concepts of interest in this paper are the notions of operational theories  and  probabilistic  models, also known as ontological models  \cite{harrigan2010einstein}. Roughly speaking, an operational theory is the list of probabilities which can be directly measured in an experiment. More precisely, any operational theory is described by a set of preparations $\mathcal{P}$, measurements $\mathcal{M}$, and probabilities $\{P(m|\mathbb{M},\mathbb{P})\}$ which determine the probability of outcome $m$ of measurement $\mathbb{M}\in\mathcal{M}$ on preparation $\mathbb{P}\in\mathcal{P}$. In general,  we can think of a preparation $\mathbb{P}$ and a measurement $\mathbb{M}$ as a list of instructions that an experimentalist follows to conduct the experiment under consideration. We assume any measurement $\mathbb{M}\in\mathcal{M}$ has a finite number of possible outcomes. 

For any set of preparations $\{\mathbb{P}_i\}_i$,  we assume their probabilistic mixtures, where one applies $\mathbb{P}_i$ with probability $p_i$ is  also a valid preparation in $\mathcal{P}$, denoted by $\mathbb{P}=\{(p_i, \mathbb{P}_i)\}$.   For instance, in quantum mechanics, a preparation $\mathbb{P}$ can be a process preparing  the ensemble $\{(p_i,\rho_i)\}$, where each density operator $\rho_i$ is prepared with probability $p_i$. Then, preparation $\mathbb{P}$ prepares the system in the density operator $\rho=\sum_i p_i \rho_i $.  Any measurement $\mathbb{M}$ in quantum mechanics is described by a POVM $\{F_m\}$, such that the probability of  outcome $m$ is given by the Born's rule, $P(m|\mathbb{M},\mathbb{P})=\Tr( \rho  F_m)$.

Given an operational theory,  we are interested in the properties of the \emph{ontological} models which explain the statistics of measurements in $\mathcal{M}$ on preparations in $\mathcal{P}$. In any such model, correlations between the choice of preparation $\mathbb{P}\in\mathcal{P}$ and the outcome $m$ of the measurement $\mathbb{M}\in\mathcal{M}$ should be mediated by  an intermediary random variable $\lambda\in \Lambda$, whose distribution is determined by preparation $\mathbb{P}$. More precisely, the probability of outcome $m$ of measurement $\mathbb{M}\in\mathcal{M}$ on preparation $\mathbb{P}\in\mathcal{P}$ is given by 
\be\label{def1}
P(m|\mathbb{M},\mathbb{P})= \sum_{\lambda\in\Lambda} \xi_{\mathbb{M}}(m|\lambda)\times \mu_{\mathbb{P}}(\lambda)  \ ,
\ee
where 
\bes\label{cond3}
\begin{align}
&\forall \lambda\in\Lambda: \mu_{\mathbb{P}}(\lambda)\ge 0\ \ ,\ \text{and}\ \  &&\sum_{\lambda\in\Lambda} \mu_{\mathbb{P}}(\lambda)=1\ , \\ 
&\forall \lambda\in\Lambda:  \xi_{\mathbb{M}}(m|\lambda) \ge 0\ ,\ &&\sum_m \xi_{\mathbb{M}}(m|\lambda) =1\ .
\end{align}
\ees
Here, each $\lambda$ is called an \emph{ontic} state,  ${\Lambda}=\{\lambda\}$  is a measurable space,  called the \emph{ontic space},  $\mu_\mathbb{P}$ is the probability distribution associated to preparation $\mathbb{P}$, and  $\xi_{\mathbb{M}}(m|\lambda)$ is the conditional probability  which defines the response of measurement  $\mathbb{M}$ for the ontic state $\lambda$. Note that the above definition and the following results hold both in the case of discrete and continuous variables, provided that $\Lambda=\{\lambda\}$ is a measurable  space. 

As usual, we assume the probabilistic model is \emph{convex-linear}, i.e.,  preparation  $\mathbb{P}=\{(p_i, \mathbb{P}_i)\}$ is described by the probability distribution 
$\mu_{\mathbb{P}}=\sum_i p_i \mu_i$, such that
\be\label{convex-lin}
\mathbb{P}=\{(p_i, \mathbb{P}_i)\}\ \ \Longrightarrow\ \ \mu_\mathbb{P}=\sum_i p_i \mu_i\ ,
\ee
 where $\mu_i$ is the probability distribution associated to preparation  $\mathbb{P}_i$  (A similar assumption is also made in the case of measurements).
  This means that to specify  $\mu_\mathbb{P}$ for a general preparation $\mathbb{P}\in\mathcal{P}$, it suffices to specify $\mu_\mathbb{P}$ for the  set of extremal  (pure) preparations, i.e., those which cannot be realized as a convex combination of other preparations.

For any operational theory, one can construct various ontological models. For instance,  as a trivial model, one can assume  the ontic state $\lambda$ uniquely determines preparation $\mathbb{P}$ (See Appendix \ref{App:A} for further discussion). In particular, in the case of a quantum mechanical system, one can consider a model whose ontic states $\{\lambda\}$ are rank-1 projectors on the Hilbert space of the system, and each pure state of the system is associated to a Dirac delta distribution. This means that the distributions associated to any pair of distinct pure states    are perfectly distinguishable, even though the pure states   themselves could have large overlaps, and hence be almost indistinguishable.  This suggests that the model is not an \emph{efficient} representation of a quantum system. 
 In particular,  a hypothetical observer who can observe the value of the ontic state $\lambda$, can send/receive an unbounded amount of information using a single qubit. However,  the  information capacity of a single qubit is bounded in quantum mechanics (In particular, the Holevo bound implies that using a single qubit one cannot transfer more than a single bit of information).  

This raises the following natural question: For any given operational theory,  what is the most \emph{economical} or most efficient ontological model? Clearly, there are various ways to formalize the notion of \emph{efficiency}. Here, we take an information theoretic approach to this problem and choose a particular measure of information which is  motivated by the notion of preparation non-contextuality.

\subsection*{Preparation Non-Contextuality}\label{Sec:Non}

 Consider two different preparations $\mathbb{P}=\{(p_i,\mathbb{P}_i)\}$ and $\mathbb{P}'=\{(p'_j,\mathbb{P}'_j)\}$  which are indistinguishable under all possible measurements, such that for any measurements $\mathbb{M}\in\mathcal{M}$ and its possible outcome $m$, the average probability of outcome $m$ is the same for both ensembles, i.e.
\be\label{eqeq}
\forall \mathbb{M}\in\mathcal{M}, \forall m: \sum_i p_i P(m|\mathbb{M},\mathbb{P}_i)=\sum_j p'_j  P(m|\mathbb{M},\mathbb{P}'_j)\ .
\ee
If this holds, we say the two preparations $\mathbb{P}$ and $\mathbb{P}'$ are \emph{operationally equivalent} and denote it by $\mathbb{P}\sim\mathbb{P}'$, or equivalently, 
\be\label{fwwgf}
\{(p_i,\mathbb{P}_i)\}\sim \{(p'_j,\mathbb{P}'_j)\}\ .
\ee
A model satisfies \emph{Preparation  Non-Contextuality (PNC)} if $\mathbb{P}\sim \mathbb{P}'$ implies that their corresponding probability distributions in the model are also equal, i.e. $\mu=\mu'$. In particular, if Eq.(\ref{fwwgf}) holds, then PNC implies
\be\label{elnk}
\sum_i p_i \mu_i=\sum_j p'_j \mu'_j\ ,
\ee
where $\mu_i$ and $\mu'_j$ are the probability distributions associated to $\mathbb{P}_i$ and $\mathbb{P}'_j$, respectively.  

For instance, for a single qubit the ensembles $\{(1/2, |0\rangle),(1/2, |1\rangle) \}$ and $\{(1/2, |+\rangle),(1/2, |-\rangle) \}$ are described by the same density operator, and hence indistinguishable under all measurements (Here, $|\pm\rangle=(|0\rangle\pm|1\rangle)/\sqrt{2}$). Therefore, preparation non-contextuality requires that 
$\mu_{0}+\mu_{1}=\mu_{+}+\mu_{-}\ ,$ where $\mu_a$ is the distribution associated to  state $|a\rangle$, for $a\in\{0,1,+,-\}$. If an operational theory does  not have a model satisfying PNC, we say the theory is preparation contextual.  It has been shown \cite{spekkens2005contextuality} that an ideal quantum mechanical system, in the absence of noise, is preparation contextual (See Appendix \ref{App:B} for a new proof). 


A fundamental question, which is the focus of this paper, is to determine if a given operational theory admits a  preparation non-contextual model. Furthermore,  for those operational theories which do not admit such a description, we  quantify the amount of deviation from this condition, i.e., the minimum amount of contextuality needed to describe the operational theory. To address these questions, we take  an information-theoretic approach.

\section{Quantifying inaccessible information}\label{Sec:Quanti}

\subsection{Definition}

Consider the  total variation distance between two probability distributions  $\mu_a$ and $\mu_b $, associated to two  preparations $\mathbb{P}_a$ and $\mathbb{P}_b$, i.e., 
\be
d_\text{TV}(\mu_a, \mu_b)\equiv \frac{1}{2}\sum_\lambda |\mu_a(\lambda)- \mu_b(\lambda)|\ .
\ee
  If this quantity is zero, then $\mathbb{P}_a \sim \mathbb{P}_b$, i.e. they are indistinguishable under all possible measurements. This follows from the monotonicity of the total variation distance under stochastic maps (data processing inequality), which  implies that for any possible measurement,  the total variation distance of the distributions of the outcomes for $\mathbb{P}_a$ and $\mathbb{P}_b$ is zero. This, in turn, implies that the distributions should be identical and hence $\mathbb{P}_a \sim \mathbb{P}_b$ (Recall that each measurement has a finite number of outcomes. Hence, the outcome distributions have zero total variation distance iff they are identical).

  Furthermore, if this  model satisfies PNC,  then the converse  also holds, i.e., 
    \be
   \mathbb{P}_a\sim  \mathbb{P}_b\ \Longleftrightarrow \ d_{\text{TV}}(\mu_a, \mu_b)=0 \ .
   \ee
    This suggests that a natural way to quantify preparation contextuality of a model is by considering the  largest distance between  distributions associated to  pairs of equivalent preparations. For an ontological model, this leads to the definition 
\be\label{kgkg}
{C}_\text{prep}\equiv \sup_{ \mathbb{P}_a \sim \mathbb{P}_b} d_\text{TV}(\mu_a,\mu_b)\ ,
\ee
 where the supremum is over all pairs of equivalent preparations  $\mathbb{P}_a, \mathbb{P}_b\in\mathcal{P}$. Note that each preparation $\mathbb{P}_a$ or $\mathbb{P}_b$ could be an ensemble $\{(p_i,\mathbb{P}_i)\}$, with an arbitrary large number of elements.  We call ${C}_\text{prep}$, which is bounded between 0 and 1, the  \emph{inaccessible information} of the model. 
 
 Clearly,  for any  model satisfying PNC, ${C}_\text{prep}=0$.  Furthermore, for any operational theory which has, at least, a pair of distinct but equivalent preparations, we can have a model with ${C}_\text{prep}=1$ (For instance, the model which associates a Dirac  delta function to any pure quantum state, has ${C}_\text{prep}=1$).  In general, finding a model which minimizes the inaccessible information can be thought of as a model selection criterion, which imposes preparation non-contextuality, if possible.

 We are interested to know if a given operational theory has a model satisfying PNC (which means ${C}_\text{prep}=0$ is achievable) and if not, what is the minimum amount of inaccessible information ${C}_\text{prep}$ needed to describe the operational theory.   To quantify this, define 
\be\label{kgkg2}
C^{\text{min}}_\text{prep}\equiv  \inf_{\text{Models}} C_{\text{prep}}=  \inf_{\text{Models}}\ \   \sup_{ \mathbb{P}_a \sim \mathbb{P}_b} d_\text{TV}(\mu_a,\mu_b)\ ,
\ee
where the infimum is taken over all ontological  models of the operational theory, i.e., over all sets of 
\be
\big(\Lambda\ ,\  \{\mu_\mathbb{P}: \mathbb{P}\in\mathcal{P} \}\ , \{\xi_{\mathbb{M}}:  \mathbb{M}\in\mathcal{M}\} \big)\ ,
\ee
 which satisfy Eqs.(\ref{def1} \ref{cond3}, \ref{convex-lin}) for the given set of probabilities $\{P(m|\mathbb{P},\mathbb{M})\}$ that define the operational theory. We call $C^{\text{min}}_\text{prep}$ the \emph{inaccessible information} of the operational theory.   By definition, this quantity  satisfies 
\be
0\le C^{\text{min}}_\text{prep}\le 1\ ,
\ee
 and, in principle,  can be anywhere in this interval. In particular, it is zero if the operational theory has a model satisfying PNC. 
  
 This quantity has a simple information-theoretic interpretation: In any model that describes the operational theory, one can find two  preparations $\mathbb{P}_a$ and  $\mathbb{P}_b$, which are indistinguishable under all possible measurements, and yet, a hypothetical observer who can observe the ontic state $\lambda$ can distinguish them with probability of success (at least) equal to  $(1+C^{\text{min}}_\text{prep})/{2}$ (assuming the two preparations are given with equal probability). Furthermore,  there exists a model for the operational theory under consideration, such that the hypothetical observer cannot distinguish two equivalent preparations with probability larger than $(1+C^{\text{min}}_\text{prep})/2$.
 
Finally, note that for any given model if the ontic space $\Lambda$ has infinite elements, then there can be two distinct distributions $\mu_a$ and $\mu_b$ with vanishing total variation distance. According to the definition of PNC in \cite{spekkens2005contextuality}, in this case the theory does not satisfy PNC, but  $C_\text{prep}$  can still be zero. However, given that the distributions with vanishing total variation distance are statistically indistinguishable, it is reasonable to slightly modify the definition of PNC in \cite{spekkens2005contextuality} to the following condition:
 \be
\mathbb{P}_a, \mathbb{P}_b\in \mathcal{P} :\ \ \ \  \mathbb{P}_a\sim \mathbb{P}_a\ \Longrightarrow\  d_{\text{TV}}(\mu_a,\mu_b)=0\ .
 \ee
Assuming this relaxation, then a model satisfies PNC iff ${C}_\text{prep}=0$.

\subsection{Inaccessible information for quantum mechanical systems}  
  
What is the inaccessible information  $C^{\text{min}}_\text{prep}$ for a quantum mechanical system? Consider the ideal case, where all pure states of the system can be prepared and all (projective) measurements can be performed.  Then, clearly, $C^{\text{min}}_\text{prep}$ can only depend on the dimension of the Hilbert space. 

To be clear, in this case the inaccessible information  of a model is defined as 
\be\label{heg}
C_\text{prep}\equiv   \sup d_{\text{TV}}\Big(\sum_i p_i \mu_i\ ,\ \sum_j p'_j \mu'_j \Big)\ ,
\ee
where the supremum is over all pairs of ensembles of preparations $\{(p_i,\mathbb{P}_i)\}$ and  $\{(p'_j,\mathbb{P}'_j)\}$  described by the same density operator, such that 
\be\label{lkhfw}
\sum_i p_i \rho_i=\sum_j p'_j \rho'_j\ ,
\ee
and $\mu_i$ and $\mu'_j$ are the probability distributions associated to preparations $\mathbb{P}_i$ and $\mathbb{P}'_j$ which prepare the system in density operators $\rho_i$ and $\rho'_j$, respectively (Note that, in general, each ensemble may have $N\rightarrow \infty$ elements, and in the limit each probability $p_i $ and $p'_j$ can go to zero).  Then,  the inaccessible information of the operational theory is defined as the $C^{\text{min}}_\text{prep}\equiv  \inf_{\text{Models}} C_{\text{prep}}$.

While finding the actual value of $C^{\text{min}}_\text{prep}$ as a function of dimension remains an open question, in this paper we show 
\begin{theorem}\label{Thm000}
For the operational theory corresponding to  a finite-dimensional  quantum system (with dimension 2 or larger) the inaccessible information $C^{\text{min}}_\text{prep}$ satisfies
\be
0.07 \approx \frac{2-\sqrt{2}}{8} \le C^{\text{min}}_\text{prep}<1  \ .
\ee
Furthermore, in the case of a single qubit, $C^{\text{min}}_\text{prep}\le 0.5$. \\
\end{theorem}

The fact that $C^{\text{min}}_\text{prep}$ is strictly larger than zero, implies that  quantum mechanics is preparation contextual, which has been known before. However,  note that the existing proofs of preparation-contextuality of quantum mechanics, do not immediately imply $C^{\text{min}}_\text{prep}>0$, because, as we discussed above, if $\Lambda$  has infinite elements then $\mu_a\neq \mu_b$ does not imply $d_\text{TV}(\mu_a,\mu_b)>0$. Therefore, $C^{\text{min}}_\text{prep}>0$ implies a stronger notion of preparation-contextuality.

  The lower bound  $\frac{2-\sqrt{2}}{8} \le C^{\text{min}}_\text{prep}$ is proven in Sec.\ref{Sec:qubit} by applying a general lower bound on $C^{\text{min}}_\text{prep}$, obtained  in theorem \ref{Thm00}, to the case of a single qubit  (See Eq.(\ref{wfkdq})). In fact, the general lower bound is expressed in terms of  experimentally measurable quantities and in the case of an ideal qubit predicts  $\frac{2-\sqrt{2}}{8} \le C^{\text{min}}_\text{prep}$. Note that by definition, the value of inaccessible information $C^{\text{min}}_\text{prep}$ for a system with a larger Hilbert space cannot be less than its value for a single qubit. Hence, this lower bound holds for any quantum mechanical system.

The general upper bound $C^{\text{min}}_\text{prep}<1$, which holds for systems with finite-dimensional Hilbert spaces,  and the special bound $C^{\text{min}}_\text{prep}\le1/2$, which holds in the case of a single qubit,  are both derived based on special ontological models, namely a general ontological model introduced by Aaronson et. al \cite{aaronson2013psi} and a model introduced by Kochen and Specker \cite{kochen1975problem}  for a single qubit. Strictly speaking, both models are defined only for pure states, but they can be easily extended to the case of mixed states as well.  Clearly, if a mixed state is prepared as an ensemble of pure states, then  its corresponding probability distribution is dictated by the convex-linearity of the model. Furthermore, if a mixed state is prepared in a different way, e.g., via purification, then the corresponding probability distribution in the model can be chosen based on a particular ensemble realization of the density operator, as a mixture of pure states.  

Then, the convexity of the total variation distance implies that  to determine  the inaccessible information  for such ontological models,  we can restrict our attention  to the ensembles of pure states. More precisely, the inaccessible information 
is determined by the total variation distance between the probability distributions associated to two \emph{ensembles of pure states} with identical density operators, i.e.  
\be\label{lejfpejf}
C_\text{prep}\equiv   \sup d_{\text{TV}}\Big(\sum_i p_i \mu_i\ ,\ \sum_j p'_j \mu'_j \Big)\ ,
\ee
where $\mu_i$ and $\mu'_j$  are the probability distributions associated to 
pure states $\psi_i$ and $\psi'_j$, and the supremum is over all pairs of ensembles $\{p_i, \psi_i\}$ and $\{p'_j, \psi'_j\}$, which satisfy 
\be\label{oijw2}
\sum_i p_i |\psi_i\rangle\langle\psi_i|=\sum_j p'_j |\psi'_j\rangle\langle\psi'_j|\ .
\ee
Assuming $C_\text{prep}$ is given by Eq.(\ref{lejfpejf}) for the ensembles of pure states satisfying Eq.(\ref{oijw2}), in Appendix \ref{App:2020} we prove that  
\begin{align}\label{kkh}
C_{\text{prep}} &\le 1- \frac{1}{2D}\Big[1-\max_{|\langle\psi'|\psi\rangle|^2\ge \frac{1}{2D}} d_{\text{TV}}(\mu,\mu')\Big] \   ,
\end{align}
where $D$ is the dimension of the Hilbert space, the maximum is over pairs of pure states $\psi$ and  $\psi'$ with $|\langle\psi'|\psi\rangle|^2\ge (2D)^{-1}$, and $\mu$ and $\mu'$ are their corresponding probability distributions in the ontological model.  

Building on a previous result of  \cite{lewis2012distinct}, 
Aaronson et. al \cite{aaronson2013psi} construct an ontological model with the property that any pair of non-orthogonal pure states $\psi$ and $\psi'$ are  described by distributions $\mu$ and $\mu'$ with a non-zero classical overlap, such that   
\be\label{vswg}
|\langle\psi|\psi'\rangle|>0\ \Longrightarrow \ d_\text{TV}(\mu,\mu')<1\ .
\ee
Therefore, for this model $\max_{|\langle\psi'|\psi\rangle|^2\ge (2D)^{-1}} d_{\text{TV}}(\mu,\mu')<1$, which by Eq.(\ref{kkh}), implies that  $C_\text{prep}$ is strictly less than one. This in turn implies the inaccessible information of the operational theory is $C^{\text{min}}_\text{prep}\equiv  \inf_{\text{Models}} C_{\text{prep}}<1$.

Next, we show that the last part of theorem \ref{Thm000}, which is a stronger upper bound on the inaccessible information of a single qubit, follows from the Kochen-Specker model of a  qubit.

\subsection{Upper bound on inaccessible information of a single qubit
via Kochen-Specker model}\label{Sec:Kochen}

In their famous work on contextuality \cite{kochen1975problem}, Kochen and Specker  also introduced a probabilistic model for a qubit.  In this model  each ontic  state is a point on the unit sphere, 
which can be denoted by the unit vector $\hat{n}\in \mathbb{R}^3$ (Equivalently, each ontic state can be thought as a pure density operator  of a qubit). Then, for any pure state $\psi$, the corresponding probability density is
\be
\mu_\psi(\hat{n})=4  \hat{n}\cdot \hat{s}_\psi \times  \Theta(\hat{n}\cdot \hat{s}_\psi)\ ,
\ee
where $ \Theta$ is the Heaviside step function and $\hat{s}_\psi$ is the Bloch vector associated to the density operator $\psi$, defined by
\be
\hat{s}_\psi=\big(\Tr(\psi\sigma_x),\Tr(\psi\sigma_y), \Tr(\psi\sigma_z)\big)\ .
\ee
Similarly, for any two-outcome projective measurement with 1-d projectors $\phi$ and $I-\phi$, the response function associated to projector $\phi$ is
\be
\xi(\phi|\hat{n})= \Theta(\hat{n}\cdot \hat{r}_\phi)\ ,
\ee
where $\hat{r}_\phi$ is the Bloch vector corresponding to 1-d projector $\phi$.  
 Kochen and Specker show that this model reproduces the Born rule, i.e. the probability of outcome corresponding to the projector $\phi$ for a measurement performed on state $\psi$ is 
\begin{align}
\int \frac{d\Omega}{4\pi}\ \mu_\psi(\hat{n}) \xi(\phi|\hat{n})= \frac{1+ \hat{r}_\phi\cdot\hat{s}_\psi}{2}= |\langle\psi|\phi\rangle|^2 \ ,
\end{align}
where $d\Omega$ is the solid angle differential. 

It can be easily seen that the Kochen-Specker model is preparation contextual \cite{leifer2013maximally} (To see this, consider two equivalent  ensembles $\{(1/2,|0\rangle),(1/2,|1\rangle)\}$ and  $\{(1/2,|+\rangle),(1/2,|-\rangle)\}$. Then, for any point on xy equator, the probability density associated to  the first ensemble vanishes, whereas for the second ensemble, the probability density is non-zero for almost all points on  this equator \cite{leifer2013maximally}). In the following, we demonstrate an upper bound on the inaccessible information for this model.

Consider two ensembles   $\{ p_i , \psi_i\}$ and  $\{ p'_j , \psi'_j\}$ described by the same density operator, such that  $\sum_i p_i \psi_i=\sum_j p'_j \psi'_j$. This implies 
\be\label{wfw}
\sum_i p_i \hat{s}_i=\sum_j p'_j \hat{s}'_j \ ,
\ee
where $\hat{s}_i$ and $\hat{s}'_j$ are the Bloch vectors of $\psi_i$ and $\psi'_j$, respectively.    Let $  \mu=\sum_i p_i \mu_i$  and $  \mu'=\sum_j p'_j \mu_j'$, where $\mu_i$ is the probability distributions associated to $\psi_i$, and is given by 
\be
\mu_i(\hat{n})=4  \hat{n}\cdot \hat{s}_i \times  \Theta(\hat{n}\cdot \hat{s}_i)\ ,
\ee
and $\mu'_j$ is the probability distributions associated to $\psi'_j$, and is defined similarly. In  Appendix \ref{App:E}, we prove that if Eq.(\ref{wfw}) holds, then
\be
d_\text{TV}(\mu,\mu')\equiv \frac{1}{2} \int \frac{d\Omega}{4\pi}\ \big|\mu(\hat{n})-\mu'(\hat{n})\big|\ \le \frac{1}{2}\ .
\ee
We conclude that for an ideal (noiselss) qubit, assuming the operational theory includes all states and all projective measurements, the inaccessible information satisfies
\be
C^\text{min}_\text{prep}\le C_\text{prep}\equiv \sup d_\text{TV}(\mu,\mu')\le \frac{1}{2}\ .
\ee
This proves the last part of theorem \ref{Thm000}.

\section{Distinguishability of Preparations }\label{Sec:dist}

In this section we introduce measures of distinguishability of preparations, which later will be used in our general lower bound on the inaccessible information in theorem \ref{Thm00}, and also in non-contextuality inequalities.   These functions are generalizations of trace distance and max-relative entropy \cite{datta2009min} in the quantum setting. Both functions are determined by  the equivalency class of preparations. Hence, although we define them in the context of operational theories,  they can also be thought of as functions in the Generalized Probabilistic Theory (GPT) \cite{hardy2001quantum, barrett2014no, schmid2019characterization} associated to the operational theory  which is obtained by quotienting relative to operational equivalences (See Sec. \ref{Sec:Discus} for further discussion). 

It is worth noting that one can consider other possible generalizations of these concepts. Since in this paper we are focused on the properties of preparations and their representation in the ontological models framework, we consider functions which are  solely determined  by the equivalency relations between preparations, as defined in Eq.(\ref{eqeq}). In particular, as it is shown in   proposition \ref{qffq}, these distinguishability measures remain invariant under finite noise.

\subsection{Operational Max-Relative Entropy}

We start by generalizing the concept of max-relative entropy \cite{datta2009min},  which itself is a generalization of $\alpha\rightarrow \infty$ limit of  R\'enyi relative entropy, defined by 
\be\label{clas}
D_\text{max}(q_a\|q_b)\equiv \log \sup_x \frac{q_a(x)}{q_b(x)}\ ,
\ee
where $q_{a,b}$ are probability distributions and $\log$ is base 2.  In quantum information theory the max-relative entropy of  a pair of density operators $\rho_a$ and $\rho_b$  is defined  \cite{datta2009min} as  
\be\label{class2}
{D}_\text{max}({\rho}_a\|\rho_b)\equiv-\log \text{max} \{y: y \rho_a \le \rho_b  \}\ ,
\ee
which reduces to the classical case in Eq.(\ref{clas}) if $\rho_a$ and $\rho_b$ commute.

Inspired by this definition,  we define the \emph{operational max-relative entropy} of a pair of preparations $\mathbb{P}_a, \mathbb{P}_b \in \mathcal{P}$ as
\begin{align}
&\mathbb{D}_\text{max}({\mathbb{P}_a}\|\mathbb{P}_b)\equiv -\log \sup y: y\le 1,  
\exists\mathbb{P}_{a'}\in\mathcal{P}: \nonumber \\ & \ \ \ \ \ \ \ \ \ \ \mathbb{P}_b\sim \{(y, \mathbb{P}_a) , (1-y, \mathbb{P}_{a'}) \}\ .
\end{align}
  In words, this means that for any  $y<2^{-\mathbb{D}_{\text{max}}({\mathbb{P}_a}\|\mathbb{P}_b)}$, there exists preparation $ \mathbb{P}_{a'}\in \mathcal{P}$, such that  $\mathbb{P}_b$ is equivalent to the  preparation  in which with probabilities $y$ and $1-y$ preparations $\mathbb{P}_a$ and $ \mathbb{P}_{a'}$ are applied. Furthermore, for $y>2^{-\mathbb{D}_\text{max}({\mathbb{P}_a}\|\mathbb{P}_b)}$ there is no preparation  $ \mathbb{P}_{a'}\in \mathcal{P}$  which satisfy this property.  Equivalently, this definition can be phrased directly in terms of the probabilities $\{P(m|\mathbb{M}, \mathbb{P})\}$ which define the operational theory:
\begin{align}
&\mathbb{D}_\text{max}({\mathbb{P}_a}\|\mathbb{P}_b)\equiv -\log \sup \big\{ y: y\le 1 , \exists \mathbb{P}_{a'} \in\mathcal{P}, \forall\mathbb{M}, m:  \nonumber \\
& y P(m|\mathbb{M}, \mathbb{P}_a)+(1-y) P(m|\mathbb{M}, \mathbb{P}_{a'})  =P(m|\mathbb{M}, \mathbb{P}_{b})
 \big\}\ . 
\end{align}

  It can be easily seen that $\mathbb{D}_{\text{max}}({\mathbb{P}_a}\|\mathbb{P}_b)=0$ if, and only if, $\mathbb{P}_a\sim \mathbb{P}_b$. Furthermore,  $\mathbb{D}_\text{max}$ is quasi-convex, i.e., for two ensembles $\mathbb{P}=\{(p_i,\mathbb{P}_i )\}$ and $\mathbb{P}'=\{(p_i,\mathbb{P}'_i )\}$, it holds that
  \be
  \mathbb{D}_\text{max}(\mathbb{P}\|\mathbb{P}')\le  \max_i  \mathbb{D}_\text{max}(\mathbb{P}_i\|\mathbb{P}_i')\ .
  \ee
  Finally, we note that $\mathbb{D}_\text{max}$ is a generalization of  max-relative entropy in Eq.(\ref{clas}) and Eq.(\ref{class2}), in the following sense:
\begin{proposition}\label{Thm-0}
Let   $\rho_a$ and $\rho_b$ be the density operators prepared by preparations $\mathbb{P}_a$ and $\mathbb{P}_b$.  If measurements in  $\mathcal{M}$  are tomographically complete, then   
\be\label{sfrfwf}
\mathbb{D}_\text{max}({\mathbb{P}_a}\|\mathbb{P}_b)\ge D_\text{max}({\rho_a}\|\rho_b)\ ,
\ee
where the  equality holds if preparations in $\mathcal{P}$ can prepare the  density operator $\tau$, which satisfies 
\be
 \rho_b=2^{-D_\text{max}(\rho_a\| \rho_b)} \rho_a+(1-2^{-D_\text{max}(\rho_a\| \rho_b)})\tau\ .
\ee 
\end{proposition}

Note that a set of measurements are called tomographically-complete if the distributions of their outcomes for a particular preparation uniquely determine the distribution of the outcomes of any other measurement for that preparation. In quantum theory,  a tomographically-complete set of measurements uniquely determines the density operator of the system.

\subsection{Operational total variation distance}\label{Sec:oper}

Next, we consider another measure of distinguishability of preparations, which is a natural generalization of the total variation distance and trace distance in the quantum setting.  Roughly speaking,  according to this measure, the distance between two preparations  $\mathbb{P}_a$ and $\mathbb{P}_b$ is the minimum amount of disturbance that should be added to each of the preparations  so that they become indistinguishable from each other. 
 Here, by disturbance we mean mixing the preparations with other preparations in $\mathcal{P}$.

Formally, for any pair of preparations $\mathbb{P}_a, \mathbb{P}_b\in\mathcal{P}$, define 
\begin{align}
&\mathbb{d}_{\text{prep}}(\mathbb{P}_a, \mathbb{P}_b)\equiv \inf_{q\ge 0} 
\frac{q}{1-q}: \exists \mathbb{P}_{a'},\mathbb{P}_{b'}\in\mathcal{P},    \nonumber \\
&\ \  \Big\{(1-q, \mathbb{P}_a), (q , \mathbb{P}_{a'})\Big\} \sim \Big\{(1-q, \mathbb{P}_b), (q, \mathbb{P}_{b'}) \Big\}\ \label{flnwfl}\ .
\end{align}
In words, $\mathbb{d}_{\text{prep}}(\mathbb{P}_a, \mathbb{P}_b)$ is the infimum of $\frac{q}{1-q}$ for $q\ge 0$, such that there exists preparations $\mathbb{P}_{a'}, \mathbb{P}_{b'}\in\mathcal{P}$ such that ensembles 
$\{(1-q, \mathbb{P}_a), (q , \mathbb{P}_{a'})\}$ and $\{(1-q, \mathbb{P}_b), (q, \mathbb{P}_{b'})\}$ are indistinguishable. Equivalently, we can directly phrase this definition in terms of probabilities $\{P(m|\mathbb{M}, \mathbb{P})\}$ that define the operational theory:
\begin{align}
&\mathbb{d}_{\text{prep}}(\mathbb{P}_a, \mathbb{P}_b)\equiv \inf  
\Big\{r\ge 0: \exists\ \mathbb{P}_{a'},\mathbb{P}_{b'}\in\mathcal{P},  \nonumber \\
&\ \ \ \ \ \ \  \forall \mathbb{M}, m:\    \frac{P(m|\mathbb{M}, \mathbb{P}_a)-P(m|\mathbb{M}, \mathbb{P}_{b})}{P(m|\mathbb{M}, \mathbb{P}_{b'})-P(m|\mathbb{M}, \mathbb{P}_{a'})}= r  \Big\}\ \label{second} . 
\end{align}
From this definition it is clear that $\mathbb{d}_{\text{prep}}(\mathbb{P}_a, \mathbb{P}_b)\le 1$.
Furthermore,  this function 
 is a metric on the space of equivalency classes of preparations, i.e.,  (i) it is non-negative,  $\mathbb{d}_{\text{prep}}(\mathbb{P}_a, \mathbb{P}_b)\ge 0$ and it is zero if, and only if,  $\mathbb{P}_a \sim \mathbb{P}_b$. (ii) It is symmetric, i.e., 
 $\mathbb{d}_{\text{prep}}(\mathbb{P}_a, \mathbb{P}_b)=\mathbb{d}_{\text{prep}}(\mathbb{P}_b, \mathbb{P}_a) $\ . (iii) As we show in Appendix \ref{App:D}, It satisfies the triangle inequality, i.e. 
 \be
 \mathbb{d}_{\text{prep}}(\mathbb{P}_a, \mathbb{P}_c)\le \mathbb{d}_{\text{prep}}(\mathbb{P}_a, \mathbb{P}_b)+\mathbb{d}_{\text{prep}}(\mathbb{P}_b, \mathbb{P}_c)\ .
\ee 
Next, we argue that $\mathbb{d}_{\text{prep}}$ generalizes the total variation distance. In fact, we show a stronger result in terms of trace distance. Recall that for any pair of density operators $\rho_a$ and $\rho_b$, their trace distance is defined as 
\be
d_\text{trace}(\rho_a,\rho_b)\equiv \frac{1}{2}\|\rho_a-\rho_b\|_1\ ,
\ee
where $\|\cdot\|_1$ is $l_1$ norm, i.e., sum of the absolute value of the eigenvlaues.  In the special case where the density operators commute with each other, trace distance reduces to the total variation distance.  Furthermore, according to Helstrom's theorem, trace distance has a simple operational interpretation: Suppose we are given a system prepared either in state $\rho_a$ or $\rho_b$ with equal probability, and the goal is to guess the given state. Then, the maximum probability of success is given by  $(1+d_\text{trace}(\rho_a,\rho_b))/2$. Moreover,  this probability of success can be achieved using the projective measurement $\{\Pi_a, \Pi_b\}$, where  $\Pi_a$ and $\Pi_b$ are, respectively, projectors to the subspaces with non-negative and negative eigenvalues of  $\rho_a-\rho_b$.

In Appendix  \ref{App:D}, we prove
\begin{proposition}\label{Thm0}
Let   $\rho_a$ and $\rho_b$ be the density operators prepared by preparations $\mathbb{P}_a$ and $\mathbb{P}_b$.  If measurements in  $\mathcal{M}$  are tomographically complete, then   
\be\label{kagg}
d_\text{trace}(\rho_a,\rho_b)\le  \mathbb{d}_\text{prep}(\mathbb{P}_a, \mathbb{P}_b) \ ,
\ee
where the  equality holds if preparations in $\mathcal{P}$ can prepare the  density operators 
\be \label{sigma}
 \sigma_{a/b}=\frac{\Pi_{a/b} (\rho_a-\rho_b) \Pi_{a/b}}{\Tr(\Pi_{a/b} (\rho_a-\rho_b))}\ ,
 \ee
where $\Pi_{a}$ and $\Pi_b$ are, respectively, projectors to the subspaces with non-negative and negative eigenvalues of  $\rho_a-\rho_b$.
\end{proposition}

\begin{figure}[htbp]
\begin{center}
\includegraphics[scale=0.57]{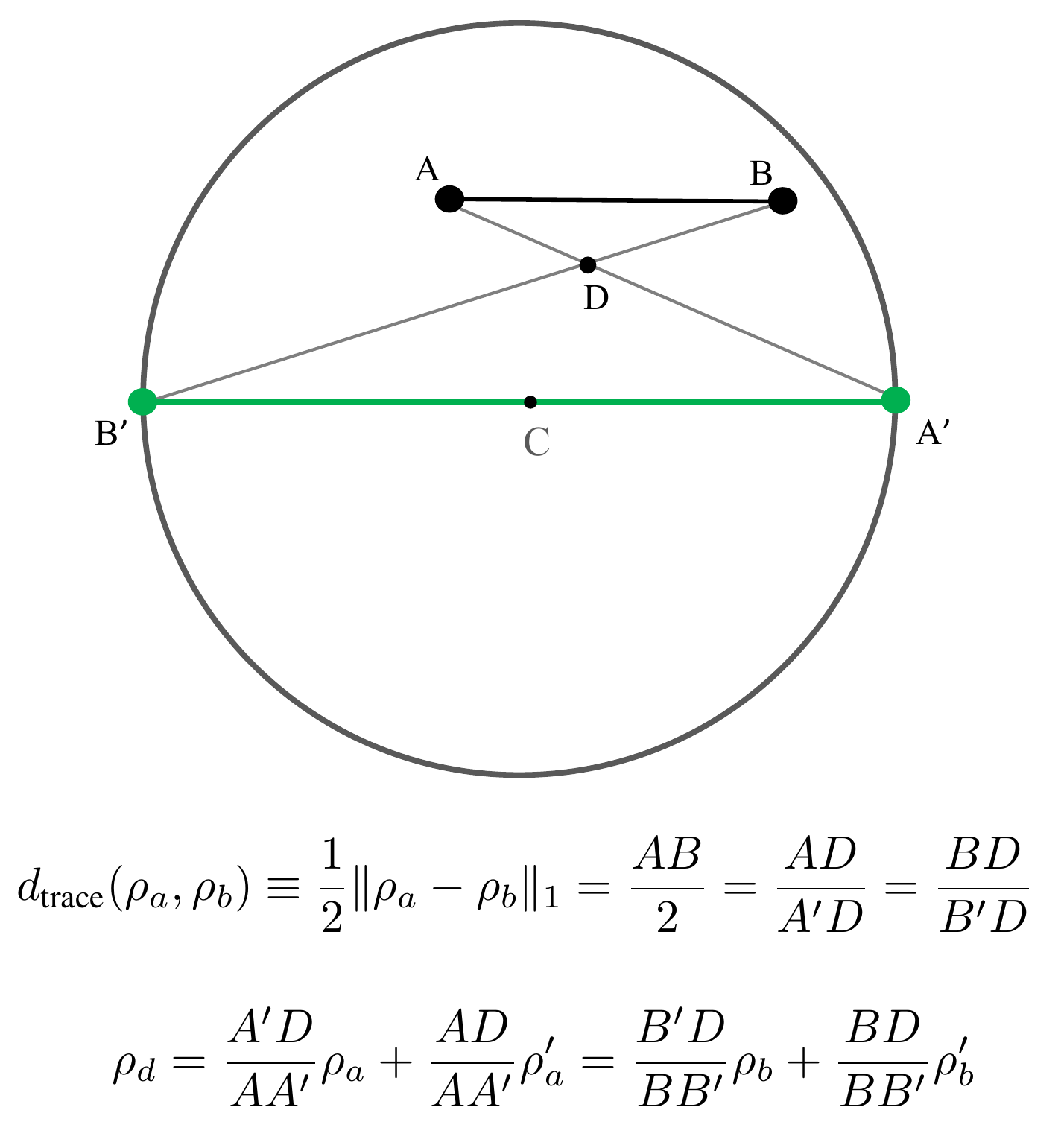}
\caption{Operational total variation distance for a qubit:
Suppose preparations $\mathbb{P}_a$ and $\mathbb{P}_b$   prepare, respectively, the density operators $\rho_a$ and $\rho_b$, which correspond to the points $A$ and $B$ on the Bloch sphere.  From definition in Eq.(\ref{second}), it follows that the operational total variation distance $\mathbb{d}_\text{prep}(\mathbb{P}_a, \mathbb{P}_b) $ is the minimum value of $r\ge 0$, for which there exists a pair of states $\rho'_{a}$ and $\rho'_b$, which can be prepared using preparations in $\mathcal{P}$ and satisfy equation $\rho_a-\rho_b=r(\rho'_b-\rho'_a)$. Let $A'$ and $B'$ be the points corresponding to $\rho'_a$ and $\rho'_b$ on the Bloch sphere. This equation implies that (i) the line segment $A'B'$ is in the same plane and parallel to the line segment $AB$, and (ii) $r=AB/A'B'$, i.e. the ratio of the Euclidian  lengths of $AB$ and $A'B'$. Therefore, to minimize $r$,  $A'$ and $B'$ should be chosen such that the length of the line segment $A'B'$ is maximized. The maximum possible length of $A'B'$ is 2 which is achieved for the diameter parallel to $AB$, denoted by  the green line $A'B'$ in the figure ($C$ is the center of sphere). We conclude that if preparations in $\mathcal{P}$ can prepare the pure states corresponding to the points $A'$ and $B'$ in this figure, then $\mathbb{d}_\text{prep}(\mathbb{P}_a, \mathbb{P}_b)=AB/A'B'=AB/2$. It can easily shown that $AB/2$, i.e. half of the Euclidian distance between $A$ and $B$,  is in fact the trace distance  of $\rho_a$ and $\rho_b$. Hence,  $\mathbb{d}_\text{prep}(\mathbb{P}_a, \mathbb{P}_b)=d_\text{trace}(\rho_a,\rho_b) $. On the other hand, if preparations in $\mathcal{P}$ cannot prepare the pure states corresponding to $A'$ and $B'$, then   $\mathbb{d}_\text{prep}(\mathbb{P}_a, \mathbb{P}_b)>d_\text{trace}(\rho_a,\rho_b) $.  It is also worth noting  that, since  the triangles $ABD$ and $A'B'D$ are similar, $AB/A'B'=AD/A'D=BD/B'D$. Since $A'B'=2$ then  $d_\text{trace}(\rho_a,\rho_b)=AB/2=AB/A'B'=AD/A'D=BD/B'D$. The intersection point $D$ corresponds to state
 $\rho_d=\frac{A'D}{AA'} \rho_a+\frac{AD}{AA'} \rho'_a =\frac{B'D}{BB'} \rho_b+\frac{BD}{BB'} \rho'_b $. This equation corresponds to the equivalency relation in Eq.(\ref{flnwfl}), which defines $\mathbb{d}_{\text{prep}}(\mathbb{P}_a, \mathbb{P}_b)$. }
\label{Bloch}
\end{center}
\end{figure}

Figure \ref{Bloch} demonstrates this result and its geometric interpretation  in the case of a single qubit.   

\subsection{The gap between the total variation distance in the model and the trace distance}

In Appendix \ref{App:D}, we show that
\begin{align}\label{dso}
d_\text{TV}(\mu_a, \mu_b) &\le ({C}_\text{prep}+1)\times \mathbb{d}_\text{prep}(\mathbb{P}_a, \mathbb{P}_b)+ {C}_\text{prep}\ ,
\end{align}
where $\mu_a$ and $\mu_b$ are the distributions associated to   preparations $\mathbb{P}_a$ and $\mathbb{P}_b$. Combining this with proposition \ref{Thm0}, we find that   in the case of a quantum mechanical system, if  measurements are tomographically complete and 
preparations in $\mathcal{P}$ can prepare density operators $\sigma_{a/b}$ in Eq.(\ref{sigma}), then 
\begin{align}
d_\text{TV}(\mu_a, \mu_b) &\le  ({C}_\text{prep}+1)\times d_\text{trace}(\rho_a,\rho_b) + {C}_\text{prep} \ .
\end{align}

Using the fact the trace distance is bounded by one, we conclude that 
\begin{theorem}\label{Thm2}
Let  $\mu_{a}$ and $\mu_b$ be the distributions associated to two preparations $\mathbb{P}_a$ and $\mathbb{P}_b$, and $\rho_{a}$ and $\rho_b$ be the corresponding density operators of the system. Then,
\begin{align}\label{koefh}
0\le  d_\text{TV}(\mu_a, \mu_b)-d_\text{trace}(\rho_a, \rho_b) &\le {C}_\text{prep}[1+d_\text{trace}(\rho_a,\rho_b)]\nonumber \\ &\le 2{C}_\text{prep}\ ,
\end{align}
where the lower bound on $d_\text{TV}(\mu_a, \mu_b)-d_\text{trace}(\rho_a, \rho_b)$ holds assuming  measurements in $\mathcal{M}$ include the projective measurement with projectors $\{\Pi_{a/b}\}$, whereas  the upper bound holds assuming measurements are tomographically complete and preparations in $\mathcal{P}$ can prepare the density operators $\sigma_{a/b}$ defined in Eq.(\ref{sigma}).
\end{theorem}
The  lower bound $0\le  d_\text{TV}(\mu_a, \mu_b)-d_\text{trace}(\rho_a, \rho_b) $, which follows  from the data processing inequality for the total variation distance, together with the Helstrom's theorem, has been also previously observed in \cite{leifer2014psi, barrett2014no}.  We also note that in the case of  an ideal quantum mechanical system, i.e. assuming all pure states can be prepared and all projective measurements can be performed, Ref. \cite{barrett2014no} and \cite{leifer2014psi} have established several upper bounds on the ratio of the classical to quantum \emph{overlaps}, i.e.,  $[1-d_\text{TV}(\mu_a, \mu_b)]/[1-d_\text{trace}(\rho_a, \rho_b)]$ (In particular, Leifer \cite{leifer2014psi} has shown that this ratio should be exponentially small in the dimension of the Hilbert space).  

As a consistency check, we note that if PNC holds, i.e.,  ${C}_\text{prep}=0$,  then theorem \ref{Thm2} implies that if $\rho_a=\rho_b$ then $\mu_a=\mu_b$, which is a restatement of  PNC. Furthermore, if PNC is violated, but the inaccessible information ${C}_\text{prep}$ is small,  then theorem implies that if $\rho_a$ and $\rho_b$   are close in the trace distance, then  their corresponding probability distributions $\mu_a$ and $\mu_b$ should also be close in the total  variation distance.  

Also, the theorem  clarifies that the implications of PNC are not limited to the special case of $\rho_a=\rho_b$. In particular, 
\begin{corollary}
If density operators $\sigma_{a/b}$ can be prepared, measurements in $\mathcal{M}$ are tomographically complete and include projective measurement $\{\Pi_{a/b}\}$, then PNC implies 
\be
d_\text{TV}(\mu_a, \mu_b)=d_\text{trace}(\rho_a, \rho_b)\ .
\ee
In the special case of pure states $\rho_{a,b}=|\psi_{a,b}\rangle\langle \psi_{a,b}|$, this means
\be
d_\text{TV}(\mu_a, \mu_b)=d_\text{trace}(\rho_a, \rho_b)=\sqrt{1-|\langle\psi_a|\psi_b\rangle|^2}\ .
\ee
\end{corollary}

Note that in this case $\sigma_{a/b}$ will be a pair of orthogonal pure states in the subspace spanned by $\{|\psi_{a}\rangle, |\psi_{b}\rangle\}$. 

It is interesting to compare this equation with a result of  Leifer and Maroney \cite{leifer2013maximally}, which shows that, under certain conditions on the set of preparations and measurements, PNC implies  $\sum_{\lambda\in \Lambda_a} \mu_b(\lambda) =|\langle\psi_a|\psi_b\rangle|^2$, where $ \Lambda_a$ is the support of $\mu_a$.

\subsection{Distinguishability of preparations in noisy quantum mechanical systems}
We saw that if preparations in $\mathcal{P}$ can prepare all quantum states of a system and measurements in $\mathcal{M}$ are tomographically complete, then the operational max-relative entropy $\mathbb{D}_\text{max}$ and  the operational  total variation distance $\mathbb{d}_\text{prep}$ reduce to, max-relative entropy $D_\text{max}$ and trace distance $d_\text{trace}$. It is also interesting to see how these quantities behave if  
preparations in $\mathcal{P}$ can only prepare noisy quantum states. More precisely, suppose the set of preparations $\mathcal{P}$ can prepare all and only states  in the form $\{\mathcal{E}(\rho)\}$ where $\rho$ is an arbitrary density operator and 
$\mathcal{E}$ is a quantum channel that describes noise on preparations. 

Suppose preparation $\mathbb{P}_i$ prepares density operator $\mathcal{E}(\rho_i)$. Assuming measurements are tomorgraphically complete, one can easily see that two ensembles  $\{(p_j, \mathbb{P}_j)\}$ and $\{(q_l, \mathbb{P}_l)\}$ are equivalent, iff 
$\sum_j p_j \mathcal{E}(\rho_j)=\sum_l q_l \mathcal{E}(\rho_l)\ $. Furthermore, assuming $\mathcal{E}$ is a one-to-one function, then this is equivalent to  
 $\sum_j p_j \rho_j=\sum_l q_l \rho_l\ $. In particular,  there exists a preparation $\mathbb{P}_c\in\mathcal{P}$, such that $\{(1-q, \mathbb{P}_a), (q, \mathbb{P}_c)\}\sim \mathbb{P}_b$ iff there exists a density operator $\rho_c$ such that $(1-q) \rho_a+q \rho_c=\rho_b$. 
 Using the definition of operational max-relative entropy, this immediately implies $\mathbb{D}_\text{max}(\mathbb{P}_a\|\mathbb{P}_b)={D}_\text{max}(\rho_a\|\rho_b)$.   We can repeat a similar argument in the case of operational total variation distance. This proves
\begin{proposition}\label{qffq}
Suppose preparations in $\mathcal{P}$ can prepare all and only  quantum states $\{\mathcal{E}(\rho)\}$, where $\rho$ is an arbitrary density operator, and $\mathcal{E}$ is a positive, trace-preserving, and one-to-one map. Consider a pair of preparations  $\mathbb{P}_a, \mathbb{P}_b\in \mathcal{P}$, which prepare states $\mathcal{E}(\rho_a)$ and $\mathcal{E}(\rho_b)$, respectively. Assuming  measurements are tomographically complete, then
\bes
\begin{align}
\mathbb{D}_\text{max}(\mathbb{P}_a\|\mathbb{P}_b)=D_\text{max}(\rho_a\|\rho_b)\ , \\ 
\mathbb{d}_\text{prep}(\mathbb{P}_a, \mathbb{P}_b)=d_\text{trace}(\rho_a,\rho_b)\ .
\end{align}
\ees
\end{proposition}
Therefore, as long as $\mathcal{E}$ remains an invertible function, the strength of noise does note affect the distinguishability of preparations, as quantified by $\mathbb{D}_\text{max}$ and $\mathbb{d}_\text{prep}$. This follows from the fact that these functions are defined solely based on the equivalency relations between preparations, which remain unchanged under a one-to-one map $\mathcal{E}$.

In Sec.\ref{Sec:noise} we use this result to determine noise thresholds for non-contextuality of quantum systems.

\section{Lower bounds on inaccessible information in terms of experimentally observable quantities}\label{Sec:Games}

In this section we  derive  lower bounds on the inaccessible information and we find a new family of non-contextuality inequalities. 


Given an operational theory with preparations $\mathcal{P}$ and measurements $\mathcal{M}$, consider a subset of preparations and measurements
\bes
\begin{align}
&\mathbb{P}_{(k, x)}\in \mathcal{P}\ ,\ \ \ \ \  &&k=1,\cdots,n; \ \  x=1,\cdots, d \\ 
&\mathbb{M}_k\in \mathcal{M}\ , \ \ \ \ \ \ &&k=1,\cdots, n\ ,
 \end{align}
 \ees
 where $n,d> 1$. For simplicity, we assume each measurement  has $d$ outcomes labeled as $y=1,\cdots, d$.  We are interested in the quantity
 \be\label{guessing}
P_\text{guess}\equiv \frac{1}{n d}\times \sum_{k=1}^n\ \sum_{x=1}^{d}  P\big(y=x \big| \mathbb{M}_k , \mathbb{P}_{(k,x)} \big)\ ,
\ee
which quantifies the correlation between the label $x$ of preparation $\mathbb{P}_{(k,x)} $ and the outcome $y$ of measurement $\mathbb{M}_k$. This quantity has a simple interpretation in terms of a guessing game:  Suppose Alice chooses an \emph{alphabet} $k\in\{1,\cdots , n\}$ and a \emph{message} $x\in \{1,\cdots, d\}$, uniformly at random, and independent of each other. Then, she applies preparation $\mathbb{P}_{(k, x)}\in \mathcal{P}$ and sends the system  to Bob.  She also reveals $k$ and asks Bob to guess $x\in\{1,\cdots, d\}$. Bob,  performs measurement  $\mathbb{M}_k\in \mathcal{M}$,  and obtains outcome $y$ which determines his guess for message $x$. He wins if $y=x$, i.e., his guess coincides with Alice's choice, which happens with probability  $P_\text{guess}$ in Eq.(\ref{guessing}).

In the following, we derive upper bounds on $P_\text{guess}$ in terms of  $C_\text{prep}^{\text{min}}$, the inaccessible information of the operational theory. First,  consider an arbitrary ontological model for this operational theory. Let $\Lambda$ be the ontic space, $\mu_{(k,x)}$ be the probability distribution associated to preparation $\mathbb{P}_{(k,x)}$ and $\xi_{\mathbb{M}_k}(x|\lambda)$ be the probability of outcome $x$ of measurement  $\mathbb{M}_k$, for the ontic state $\lambda$.  Using Eq.(\ref{def1}) and Eq.(\ref{cond3}), we find
\bes\label{wgg}
 \begin{align}
P_\text{guess}&=  \frac{1}{n} \sum_k \frac{1}{d}\sum_{x=1}^{d}  \sum_\lambda \xi_{\mathbb{M}_k}(x|\lambda) \mu_{(k,x)}(\lambda) \label{krk}\\ 
&\le  \frac{1}{d}  \sum_\lambda  \frac{1}{n} \sum_k  \max_{x}\mu_{(k,x)}(\lambda) \\ &= \frac{1}{d}  \sum_{\lambda} \max_{\textbf{x}} \mu_{\textbf{x}}(\lambda)\ ,
 \end{align}
 \ees
where $\textbf{x}=x_1,\cdots, x_n\in\{1,\cdots,d\}^n$ , and
\be\label{Eq2020}
\mu_{\textbf{x}}\equiv \frac{1}{n}\sum_{k} \mu_{(k,x_k)}\ .
\ee
Note that $\mu_{\textbf{x}}$ is the probability distribution associated to the ensemble $\mathbb{P}_\textbf{x}=\{(1/n,\mathbb{P}_{(k,x_k)} ): k=1,\cdots, n\}$, i.e., the preparation in which with probability $1/n$ one applies preparation $\mathbb{P}_{(k,x_k)}$.

 Similarly, 
\begin{align}
&\sum_{x=1}^{d} \mu_{(k,x)}(\lambda)\ge \max_{x}\mu_{(k,x)}(\lambda)+ (d-1) \min_{x}\mu_{(k,x)}(\lambda) \nonumber\\ 
&\ge \sum_{x=1}^{d}   \xi_{\mathbb{M}_k}(x|\lambda) \mu_{(k,x)}(\lambda)+ (d-1) \min_{x}\mu_{(k,x)}(\lambda)\ ,
\end{align}
where the first line follows from the fact that there are $d$ terms in the summation and the second line follows from  Eq.(\ref{cond3}). Together with Eq.(\ref{krk}), this implies $P_\text{guess}\le 1-\frac{d-1}{d}  \sum_{\lambda} \min_{\textbf{x}} \mu_{\textbf{x}}(\lambda)$. To summarize, we find
 \begin{align}\label{wfjej}
P_\text{guess}\le \min\Big\{\frac{1}{d}  \sum_{\lambda} \max_{\textbf{x}} \mu_{\textbf{x}}(\lambda), 1-\frac{d-1}{d}  \sum_{\lambda} \min_{\textbf{x}} \mu_{\textbf{x}}(\lambda)  \Big\}\ .
 \end{align}

In Appendix \ref{App:lemma}, we prove the following lemma, which puts an upper bound on the right-hand side of Eq.(\ref{wfjej}) in terms of $C_\text{prep}$, the inaccessible information of the model,  together with quantities which can be directly determined from the operational theory (i.e., can be estimated from the experimental data).   
\begin{lemma}\label{lem}
Let $\mu_z$ be the probability associated to preparation $\mathbb{P}_z\in \mathcal{P}$, where $z=1,\cdots , N$. Then, 
\bes
\begin{align}
\sum_{\lambda} \max_z \mu_z(\lambda) &\le \gamma_1 (1 + N\times C_\text{prep})\ , \\
\sum_\lambda \min_{z}  \mu_{z}(\lambda)&\ge \gamma_2^{-1}-N \times  C_\text{prep}\ ,
\end{align}
\ees
where $C_\text{prep}\equiv \sup_{ \mathbb{P}_a \sim \mathbb{P}_b} d_\text{TV}(\mu_a,\mu_b)$ is the  inaccessible information of the  model,  
$\gamma_1 \equiv \inf_{\mathbb{P}_f\in\mathcal{\mathcal{P}}}\ \max_{z}  2^{\mathbb{D}_\text{max}(\mathbb{P}_z \|\mathbb{P}_f)}$, and $ \gamma_2 \equiv \inf_{\mathbb{P}_f\in\mathcal{\mathcal{P}}}\ \max_{z}  2^{\mathbb{D}_\text{max}(\mathbb{P}_f \|\mathbb{P}_z)}\ .$
\end{lemma}

\subsection{Main result I: Experimentally measurable lower bounds on the inaccessible information }

Combining this lemma with Eq.(\ref{wfjej}) and taking the infimum of $C_\text{prep}$  over all models, we find 
\begin{theorem}\label{Thm00}
The guessing probability defined in Eq.(\ref{guessing}) is upper bounded by
\bes\label{ehre}
\begin{align}
P_\text{guess} &\le \frac{\alpha_\text{min}}{d}(1+C^\text{min}_\text{prep} \times d^n)\ ,\\
P_\text{guess} &\le 1-\frac{d-1}{d} \beta^{-1}_\text{min} + (d-1)d^{n-1} \times  C^\text{min}_\text{prep}\label{wf11}\ ,
\end{align}
\ees
where  $C^{\text{min}}_\text{prep}\equiv   \inf_{\text{Models}}\ \   \sup_{ \mathbb{P}_a \sim \mathbb{P}_b} d_\text{TV}(\mu_a,\mu_b)\ $ is the inaccessible information of the operational theory, as defined in Eq.(\ref{kgkg2}), and
\bes\label{wwfwf}
\begin{align}
\alpha_\text{min}&\equiv {\inf_{\mathbb{P}_f\in\mathcal{\mathcal{P}}}}\ \max_{\bf{x}}  2^{\mathbb{D}_\text{max}(\mathbb{P}_{\bf{x}} \|\mathbb{P}_f)}\  ,\\
\beta_\text{min}&\equiv \inf_{\mathbb{P}_f\in\mathcal{\mathcal{P}}}\ \max_{\bf{x}}  2^{\mathbb{D}_\text{max}(\mathbb{P}_f\|\mathbb{P}_{\bf{x}})}\ .
\end{align}
\ees
\end{theorem}

These bounds immediately yield lower bounds on the inaccessible information of the operational theory:  
\bes\label{aefk}
\begin{align}
C^{\text{min}}_\text{prep} &\ge \frac{1}{d^n}\Big[\frac{d\times P_\text{guess}}{\alpha_\text{min}}-1\Big]\ , \\ 
C^\text{min}_\text{prep}&\ge \frac{P_\text{guess}-(1-\frac{d-1}{d} \beta^{-1}_\text{min} )}{(d-1)d^{n-1}}  \ .
\end{align}
\ees

Using this result, we can experimentally demonstrate a lower bound on $C^\text{min}_\text{prep} $. To achieve this we need to (i) measure $P_\text{guess}$ defined in Eq.(\ref{guessing}), for a properly chosen set of preparations $\{\mathbb{P}_{(k,x)}\}$ and measurements $\{\mathbb{M}_{k}\}$, and (ii) choose a   tomographically complete  set of measurements $\mathcal{M}$ and measure them for a set of preparations $\mathcal{P}$, which includes all preparations $\{\mathbb{P}_{(k,x)}\}$.  This gives  a list of probabilities $\{P(m|\mathbb{M},\mathbb{P})\}$ which define  the operational theory. Having this list, we can  immediately  calculate  $\alpha_\text{min}$ and $\beta_\text{min}$ defined in Eq.(\ref{wwfwf}). Then,  applying the above results, we obtain a lower bound on $C^{\text{min}}_\text{prep}$. 
 Note that  if both $\mathcal{P}$ and $\mathcal{M}$  contain, respectively, a finite set of preparations and measurements, together with their probabilistic mixtures, then to determine parameters  $\alpha_\text{min}$ and $\beta_\text{min}$,  we only need to find a finite list of probabilities $\{P(m|\mathbb{M},\mathbb{P})\}$.

Also, note that, in general,  the values of $\alpha_\text{min}$ and $\beta_\text{min}$  depend on the choice of the set of preparations $\mathcal{P}$. In particular, by adding more preparations to this set, we may reduce these quantities, which results in  stronger lower bounds on  $C^\text{min}_\text{prep} $.  In Sec.(\ref{Sec:qq}) we determine the lowest possible values of $\alpha_\text{min}$ and $\beta_\text{min}$ in the quantum setting, as well as the smallest set of preparations $\mathcal{P}$ which allows us to achieve  these minimum values.

\subsection{A new class of robust non-contextuality inequalities}

Next, we consider the special case of $C^{\text{min}}_\text{prep}=0$, i.e., when the  inaccessible information of the operational theory is zero.  
This corresponds to the case where the operational theory is preparation non-contextual, i.e., can be described by a model satisfying PNC. In this case, Eq.(\ref{ehre}) implies
\be\label{wfwfw}
P_\text{guess}\le \min\{\frac{\alpha_\text{min}}{d}, 1-\frac{d-1}{d} \beta^{-1}_\text{min}\}\ .
\ee

Hence, to experimentally demonstrate  that quantum mechanics is preparation contextual, it suffices to show violation of this inequality, which is analogous to the experimental  violation of Bell's inequality.  As another application, in Sec.\ref{Sec:noise} we use this inequality to find a noise threshold for preparation  contextuality. 

 Note that the quantities $\alpha_\text{min}$ and $\beta_\text{min}$ only depend on the operational equivalencies in the operational theory. This type of bounds, which put constraints on the operational theory based on (i) the assumption of PNC and (ii)  the operational equivalencies,  are called  non-contextuality inequalities (See e.g. \cite{mazurek2016experimental, schmid2018contextual, schmid2018all}).  In fact, as we see in Sec.\ref{Sec:qubit}, a previously known non-contextuality inequality is a special case of Eq.(\ref{wfwfw}). 

A nice feature of the lower bounds on the inaccessible information and the resulting non-contextuality inequalities in Eq.(\ref{wfwfw})  is their robustness against imperfections in experiments. Recall that the definition of preparation non-contextuality is based on the existence of distinct, but equivalent preparations, such that for any measurements the statistics of the outcomes on the two preparations are indistinguishable (In quantum mechanics, this is the case, for instance, for two ensembles $\{(1/2,|0\rangle),(1/2,|1\rangle)\}$ and $\{(1/2,|+\rangle),(1/2,|-\rangle)\}$).

However, in actual experiments, due to various  errors, preparing different  ensembles with exactly identical density operators is impossible. Hence, it may not be clear how one can experimentally study the consequences of preparation non-contextuality. To address this issue, \cite{pusey2018robust} and \cite{mazurek2016experimental} have developed a technique for forming equivalent preparations by considering mixtures of inequivalent preparations.

Remarkably, our non-contextuality inequality in Eq.(\ref{wfwfw}) does not suffer from this issue, because  when one calculates the parameters $\alpha_\text{min}$ and $\beta_\text{min}$ from the experimental data $\{P(m|\mathbb{M},\mathbb{P} )\}$,  this process automatically finds certain pairs of equivalent preparations, which can be obtained by mixing the actual preparations  realized in the experiment.

\subsection{Tightness of the bound in the classical case}

To understand this non-contextuality inequality better and show its tightness, we consider the following example, which can be understood independently of the above results:  Suppose Alice randomly chooses one of the distributions $\{\mu_{k,x}: k=1,\cdots n; x=1,\cdots d\}$ uniformly at random, i.e., each with probability $(dn)^{-1}$. Then, she  generates a sample $\lambda\in\Lambda$ with probability $\mu_{(k,x)}(\lambda)$, and informs Bob about the values of $\lambda$ and $k$. Bob should guess the value of $x$. 

It can be easily seen that Bob's optimal strategy is to guess the value  $x$ which maximizes the probability $\mu_{(k,x)}(\lambda)$, for the given values of $k$ and $\lambda$. 
Then, given a particular value of $k$, he succeeds with probability $d^{-1}\times \sum_\lambda \max_{x} \mu_{(k,x)}(\lambda)$.  Therefore, the maximum achievable  guessing probability in this case is
\bes\label{kskks}
\begin{align}
P^{\text{max}}_\text{guess} &=\frac{1}{n} \sum_{k=1}^n  \frac{1}{d} \sum_\lambda \max_{x} \mu_{(k,x)}(\lambda)\\ &=\frac{1}{d}  \sum_\lambda  \max_{{\bf{x}}} \mu_{\bf{x}}(\lambda)\ ,
\end{align}
\ees
where $\mu_{\bf{x}}=n^{-1}\sum_k \mu_{(k,x_k)}$ and the maximum is over  $\textbf{x}=x_1\cdots x_n\in\{1,\cdots, d\}^n$. 

For any positive function $f:\Lambda\rightarrow \mathbb{R}_{\ge 0}$, it can be easily seen that
\be
\sum_\lambda f(\lambda)=\inf_p \sup_\lambda \frac{f(\lambda)}{p(\lambda)}=\sup_\lambda \frac{f(\lambda)}{p_\ast(\lambda)}\ ,
\ee
where the infimum  is over all probability distributions over $\Lambda$, and $p_\ast(\lambda)\equiv f(\lambda)\times[\sum_{\lambda'} f(\lambda')]^{-1}$. Combining this fact with  the definition of the max-relative entropy for probability distributions $\mu$ and $\nu$,  i.e. 
\be
2^{D_\text{max}(\mu\|\nu)}\equiv\sup_\lambda \frac{\mu(\lambda)}{\nu(\lambda)}\ ,
\ee
we obtain
\begin{align}\label{clas2}
  \sum_\lambda  \max_{{\bf{x}}} \mu_{\bf{x}}(\lambda)&= \inf_{\nu}  \max_{\textbf{x}} 2^{D_\text{max}(\mu_{\bf{x}}\|\nu)}= \max_{\textbf{x}} 2^{D_\text{max}(\mu_{\bf{x}}\|\nu_\ast)}\ ,
\end{align}
where the infimum is over all probability distributions on $\Lambda$, and 
\be
\nu_\ast(\lambda)\equiv\frac{\max_{\textbf{x}} \mu_\textbf{x}(\lambda) }{\sum_{\lambda'} \max_{\textbf{x}} \mu_\textbf{x}(\lambda')}\ .
\ee
Using Eq.(\ref{kskks}), it follows that
\begin{align}
P^{\text{max}}_\text{guess} &=\frac{1}{d}\inf_{\nu}  \max_{\bf{x}} 2^{D_\text{max}(\mu_{\bf{x}}\|\nu)}= \frac{1}{d}\ \max_{\bf{x}} 2^{D_\text{max}(\mu_{\bf{x}}\|\nu_\ast)}\  .
\end{align}
To compare this result with  our general bounds on the guessing probability in Eq.(\ref{wfwfw}), we describe the above game as an operational theory. In this operational theory, preparation $\mathbb{P}_{(k,x)}$ prepares the system in distribution  $\mu_{(k,x)}$, and there exists a measurement which determines the value of the ontic  state $\lambda\in\Lambda$ with certainty. Clearly, for this operational theory $C_\text{prep}^{\text{min}}=0$.   Furthermore,
\bes
 \begin{align}
 \alpha_\text{min}&\equiv {\inf_{\mathbb{P}_f\in\mathcal{\mathcal{P}}}}\ \max_{\bf{x}}  2^{\mathbb{D}_\text{max}(\mathbb{P}_{\bf{x}} \|\mathbb{P}_f)}\\ &\ge  \inf_{\nu}   \max_{\bf{x}} 2^{D_\text{max}(\mu_{\bf{x}}\|\nu)}\\ &=  \max_{\bf{x}} 2^{D_\text{max}(\mu_{\bf{x}}\|\nu_\ast)}\ , 
  \end{align}
\ees
 and the inequality holds as equality if preparations in $\mathcal{P}$ can prepare the distribution $\nu_\ast$.

 We conclude that if there exists a measurement determining the value of the ontic  state with certainty, and if there is a preparation whose corresponding probability distribution is $\nu_\ast$, then 
 \begin{align}
P^{\text{max}}_\text{guess} &=\frac{\alpha_\text{min}}{d}\  ,
\end{align}
 which means our non-contextuality inequality $P_\text{guess} \le \frac{\alpha_\text{min}}{d}$ holds as equality. 
 
\section{Inaccessible information in quantum mechanical systems}
\label{Sec:qq}
In this section, we show that quantum mechanics predicts that the above non-contextuality inequalities can be violated and the inaccessible information  $C^{\text{min}}_\text{prep}$ is non-zero in certain experiments. We start by determining  the quantities $\alpha_\text{min}$ and $ \beta_\text{min}$ in the quantum setting.

Let $\rho_{(k,x)}$ be the density operator prepared by $\mathbb{P}_{(k,x)}$. Then, from  proposition \ref{Thm-0}, we can easily see that  if measurements in $\mathcal{M}$ are tomographically complete, then
\bes\label{wehhw}
\begin{align}
 \alpha_\text{min} &\equiv  \inf_{\mathbb{P}_f\in\mathcal{\mathcal{P}}}\ \max_{\i}  2^{\mathbb{D}_\text{max}(\mathbb{P}_{\textbf{x}} \|\mathbb{P}_f)} \ge \inf_{\sigma_f} \max_{\textbf{x}} 2^{D_\m(\rho_{\textbf{x}}\|\sigma_f)}\ \label{affq} ,\\
 \beta_\text{min} &\equiv  \inf_{\mathbb{P}_f\in\mathcal{\mathcal{P}}}\ \max_{\i}  2^{\mathbb{D}_\text{max}(\mathbb{P}_f\|\mathbb{P}_{\textbf{x}}) } \ge \inf_{\sigma_f} \max_{\textbf{x}} 2^{D_\m(\sigma_f\|\rho_{\textbf{x}})}\ ,
\end{align}
 \ees
 where  $\rho_{\textbf{x}}\equiv \frac{1}{n} \sum_{k=1}^n \rho_{(k,x_k)}$, i.e.,  the density operator prepared by $\mathbb{P}_{\textbf{x}} $,  and  $\inf_{\sigma_f}$ is the infimum over the set of all density operators of the system. In general, if preparations in $\mathcal{P}$ cannot prepare all density operators of the system, then $\alpha_\text{min}$ and $\beta_\text{min}$ could be strictly larger than the lower bounds in Eqs.(\ref{wehhw}), which makes the lower bounds on the accessible information  weaker.   However, using  proposition \ref{Thm-0}, we can easily see that   there exists a finite set of states $\{\tau_\textbf{x}\}_{\textbf{x}}$ such that if preparations in $\mathcal{P}$ can prepare all of them, then Eqs.(\ref{wehhw}) hold as equality. In Sec.\ref{Sec:qubit} we will discuss several  qubit examples.

 \subsection{Interpreting the non-contextuality inequality in terms of two variants of the guessing game}\label{Sec:lkhf}
 
In the quantum setting, assuming all POVM measurements are possible, the maximum guessing probability  can be expressed in terms of the max-relative entropy. This follows from the result of \cite{datta2009min, konig2009operational, mosonyi2009generalized}: Suppose one is given a quantum system  in the density operator $\rho_{l}$, where $l\in\{1,\cdots, L\}$ is chosen uniformly at random.  Then, the maximum achievable  probability of guessing the correct label $l\in\{1,\cdots , L\}$  is 
 \begin{align}\label{QM}
\max_{\{B_l\}} \frac{1}{L}\sum_l  \Tr(B_l \rho_{l})= \frac{1}{L} \inf_\sigma \max_{l\in\{1,\cdots, L\}} 2^{D_\m(\rho_{l}\|\sigma)}\ ,
\end{align} 
where the maximum  is over all POVM's, and the infimum  is over all density operators of the system \cite{datta2009min, konig2009operational, mosonyi2009generalized} (Note that this equality can be thought as a generalization of the first equality in Eq.(\ref{clas2}).  Using this result, we can determine the maximum guessing probability $P_\text{guess}$ in Eq.(\ref{guessing}) for quantum mechanical systems. Also, as we show next, this result reveals an interesting interpretation  of the non-contextuality bound $P_\text{guess}\le \alpha_\text{min}/d$.

This interpretation is based on a modified version of the guessing game in which Alice does not reveal  the  alphabet $k\in \{1,\cdots, n\}$ to Bob, but  she allows him to return  a  string $\textbf{y}=y_1y_2\cdots y_n$, where $y_k\in \{1,\cdots , d\}$ is Bob's guess corresponding to alphabet $k$. He wins if $y_k=x$, i.e., if his guess for the case where the alphabet is $k$, coincides with Alice's choice of message $x$.  In this case, since he does not know $k$, Bob performs a fixed measurement $\mathbb{M}$  and  wins with probability
 \bes\label{guessing2}
  \begin{align}
Q_\text{guess}&\equiv \frac{1}{d\times  n}  \sum_{x=1}^{d}\sum_{k=1}^n \sum_{\textbf{y}}  \delta_{y_k, x}  P(\textbf{y}|\mathbb{M}, \mathbb{P}_{(k,x)})\\  &=\frac{1}{d} \sum_{\textbf{x}}  P(\textbf{x}|\mathbb{M}, \mathbb{P}_{\textbf{x}})=d^{n-1} \times   R_\text{guess}\ , 
\end{align}
\ees
where $\mathbb{P}_{\textbf{x}}$, defined below Eq.(\ref{Eq2020}), is the preparation process where one applies preparations $\mathbb{P}_{(k,x_k)}: k=1,\cdots, n$, each with probability $1/n$, and   $ R_\text{guess}$ is the guessing probability in the game where Alice  chooses each $\textbf{x}\in\{1,\cdots, d\}^n$ uniformly at random, i.e.,  with probability $d^{-n}$, then applies preparation $\mathbb{P}_{\textbf{x}}$ and sends the system to Bob. Bob performs a measurement $\mathbb{M}$ and wins if he guesses the string $\textbf{x}$ correctly.

Suppose preparations in $\mathcal{P}$ can prepare all states $\{\tau_\textbf{x}\}$, which are needed to have the equality in Eq.(\ref{affq}).  If this assumption is satisfied, then  combining Eq.(\ref{QM}) and Eq.(\ref{guessing2}), we find that the maximum guessing probability in the modified game, where Bob is given state $\rho_{(k,x)}$, but he does not know the value of the alphabet $k$, is given by
\bes\label{ggjf}
 \begin{align}
Q^{\text{QM}}_\text{guess} &\equiv  \max_{\{B_\textbf{y}\}}   \frac{1}{d\times  n}  \sum_{x=1}^{d}\sum_{k=1}^n \sum_{\textbf{y}}  \delta_{y_k, x}  \Tr(\rho_{(k,x)} B_\textbf{y})\\ &= \frac{1}{d}  \max_{\{B_\textbf{y}\}} \sum_{\textbf{y}}  \Tr(\rho_\textbf{y} B_\textbf{y})\\ &=\frac{1}{d}\inf_{\sigma_f} \max_{\textbf{x}} 2^{D_\m(\rho_{\textbf{x}}\|\sigma_f)}\\ &=  \frac{\alpha_\text{min}}{d}\ ,
\end{align}
\ees
where the maximums in the first and second lines are over all possible POVM's.  Therefore, if Eq.(\ref{affq}) holds as equality, then the non-contextuality inequality $P_\text{guess}\le\alpha_\text{min}/d$  can be interpreted as
 \be
  P_\text{guess}\le \frac{\alpha_\text{min}}{d}=Q^{\text{QM}}_\text{guess} \ ,
 \ee
and the lower bound on $C^{\text{min}}_\text{prep}$ in Eq.(\ref{aefk}) can be rewritten as
\be\label{kjw0}
C^{\text{min}}_\text{prep} \ge \frac{1}{d^n}\Big[\frac{P_\text{guess}}{Q^{\text{QM}}_\text{guess}}-1\Big] \ .
\ee

This means that if $P_\text{guess}$ is strictly larger than $Q^{\text{QM}}_\text{guess}$, which means  knowing the alphabet $k$ gives Bob an advantage for guessing the message $x$, then  $C^{\text{min}}_\text{prep}>0$, and therefore we have a proof of preparation contextuality of quantum mechanics.  For instance, suppose alphabet $k$ determines the basis in which the information about message $x$ is encoded. Then, due to the information-disturbance principle,  without knowing the basis,  Bob's success probability in guessing the encoded message $x$ is reduced, which means $Q^{\text{QM}}_\text{guess}$ is strictly  less than $  P_\text{guess}$. 

\subsection{Qubit Case}\label{Sec:qubit}
Next, we  consider several qubit examples.  We restrict our attention to the special case of  $d=2$, i.e., when Bob should perform binary measurements.  Furthermore, to simplify the discussion, assume the uniform mixture of states $\{\rho_{(k,x)}\}$ is the maximally mixed state, i.e.  $\frac{1}{2n} \sum_k \sum_x \rho_{(k,x)}=\frac{I}{2}$. Let $ \vec{n}_{(k,x_k)}$ be the Bloch vector corresponding to $\rho_{(k,x)}$, and  for any string $\textbf{x}\in\{1,2\}^n$, define
\be
\vec{n}_\textbf{x}=\frac{1}{n} \sum_{k=1}^n \vec{n}_{(k,x_k)} \ ,
\ee
i.e., the Bloch vector corresponding to $\rho_\textbf{x}=n^{-1}\sum_{k} \rho_{(k,x_k)}$. 

Assume preparations in $\mathcal{P}$ can  prepare all states  which are needed to achieve the equality in  Eqs.(\ref{wehhw}) (We specify these states below). If Eqs.(\ref{wehhw}) hold as equality, then
 \bes\label{efef}
 \begin{align}
\alpha_\text{min} &\equiv \inf_{\mathbb{P}_f\in\mathcal{\mathcal{P}}}\ \max_{\textbf{x}}  2^{\mathbb{D}_\text{max}(\mathbb{P}_{\textbf{x}}\|\mathbb{P}_f )}\\ &=\inf_{\sigma}\ \max_{\textbf{x}}  2^{D_\text{max}(\rho_\textbf{x}\|\sigma)} \label{kgk}\\ &=\max_{\textbf{x}}  2^{D_\text{max}(\rho_\textbf{x}\|\frac{I}{2})}\\  &= 1+ \max_{\textbf{x}} \|n_\textbf{x}\|\  .
\end{align}
\ees
Here, to get the third line we have used the fact that in the second line the infimum is achieved for $\sigma=I/2$. This follows from the assumption that $\frac{1}{2n} \sum_k \sum_x \rho_{(k,x)}=\frac{I}{2}$ together with the fact that $D_\text{max}$ is a quasi-convex function. Also, the last line follows from the fact that $\rho_\textbf{x}$ and $\frac{I}{2}$ commute with each other, and therefore $2^{D_\text{max}(\rho_\textbf{x}\|\frac{I}{2})}$ is the maximum ratio of the eigenvalues of  $\rho_\textbf{x}$, i.e. $(1\pm \|n_\textbf{x}\|)/2 $,  to the corresponding eigenvalue for $I/2$. Similarly, we can easily show that
\bes\label{efef2}
\begin{align}
\beta_\text{min} &\equiv \inf_{\mathbb{P}_f\in\mathcal{\mathcal{P}}}\ \max_{\textbf{x}}  2^{\mathbb{D}_\text{max}(\mathbb{P}_f\|\mathbb{P}_{\textbf{x}})}\\ &=\inf_{\sigma}\ \max_{\textbf{x}}  2^{D_\text{max}(\sigma\|\rho_\textbf{x})} \\  &=\frac{1}{1- \max_{\textbf{x}} \|\vec{n}_\textbf{x}\|}\ \ .
\end{align}
\ees
Using the fact that in both cases the infimums are achieved for $\sigma=I/2$, we can easily see that to achieve equality in Eqs.(\ref{wehhw}),  preparations in $\mathcal{P}$ need to  prepare states
 \be\label{wfqe}
 \tau_{\textbf{x},\pm}=\frac{1}{2}(I\pm \frac{\vec{n}_{\textbf{x}}}{ \|\vec{n}_{\textbf{x}}\|}\cdot \vec{\sigma}) ,\ \ \forall\textbf{x}\in \{1,2\}^n\ .
 \ee
In particular, if $\mathcal{P}$ prepares states $\{ \tau_{\textbf{x},+}\ , \textbf{x}\in\{1,\cdots, d\}^n\}$, then $\beta_\text{min} =(1- \max_{\textbf{x}} \|\vec{n}_\textbf{x}\|)^{-1}$, and if it prepares all states $\{ \tau_{\textbf{x},-}\ , \textbf{x}\in\{1,\cdots, d\}^n\}$, then $\alpha_\text{min}= 1+ \max_{\textbf{x}} \|n_\textbf{x}\|$.

Interestingly, in this case we find that
\be
1-\frac{d-1}{d}\beta^{-1}_\text{min}=\frac{\alpha_\text{min}}{d}=Q^{\text{QM}}_\text{guess}\ ,
\ee
which means the two non-contextuality inequalities in Eq.(\ref{wfwfw}) in terms of  $\alpha_\text{min}$ and $\beta_\text{min}$ coincide.

Finally, using the second bound in Eq.(\ref{aefk}), i.e. the bound in terms of $\beta_\text{min}$, we find 
\begin{align}\label{bound99}
C^\text{min}_\text{prep}&\ge \frac{P_\text{guess}-Q^{\text{QM}}_\text{guess}}{2^{n-1}}  \ .
\end{align}
It turns out that this bound is stronger than the bound in Eq.(\ref{kjw0}), which is obtained based on $\alpha_\text{min}$.

\subsubsection*{Examples} 

Suppose the pair of states corresponding to the same alphabet $k$, are orthogonal pure states. Then, ideally it is possible to achieve $P_\text{guess}=1$.  Furthermore, because orthogonal states are represented by opposite points  on the Bloch sphere, to  maximize $\|\vec{n}_\textbf{x}\|=\|\sum_{k=1}^n \vec{n}_{(k,x_k)}\|/n$,  the string $\textbf{x}=x_1\cdots x_n \in\{1,2\}^n$  should be chosen such that the Bloch vectors $\vec{n}_{(k,x_k)}$ are all in the same hemisphere, namely the hemisphere in which all vectors have non-negative components in the direction of the average vector $\vec{n}_{\textbf{x}}$. In other words, the problem of finding $\textbf{x}$ which maximizes $\|\vec{n}_{\textbf{x}}\|$ is equivalent to finding the hemisphere for which the length of the average Bloch vector for vectors  inside that hemisphere is maximized.

\begin{figure}[htbp]
\begin{center}
\includegraphics[scale=0.5]{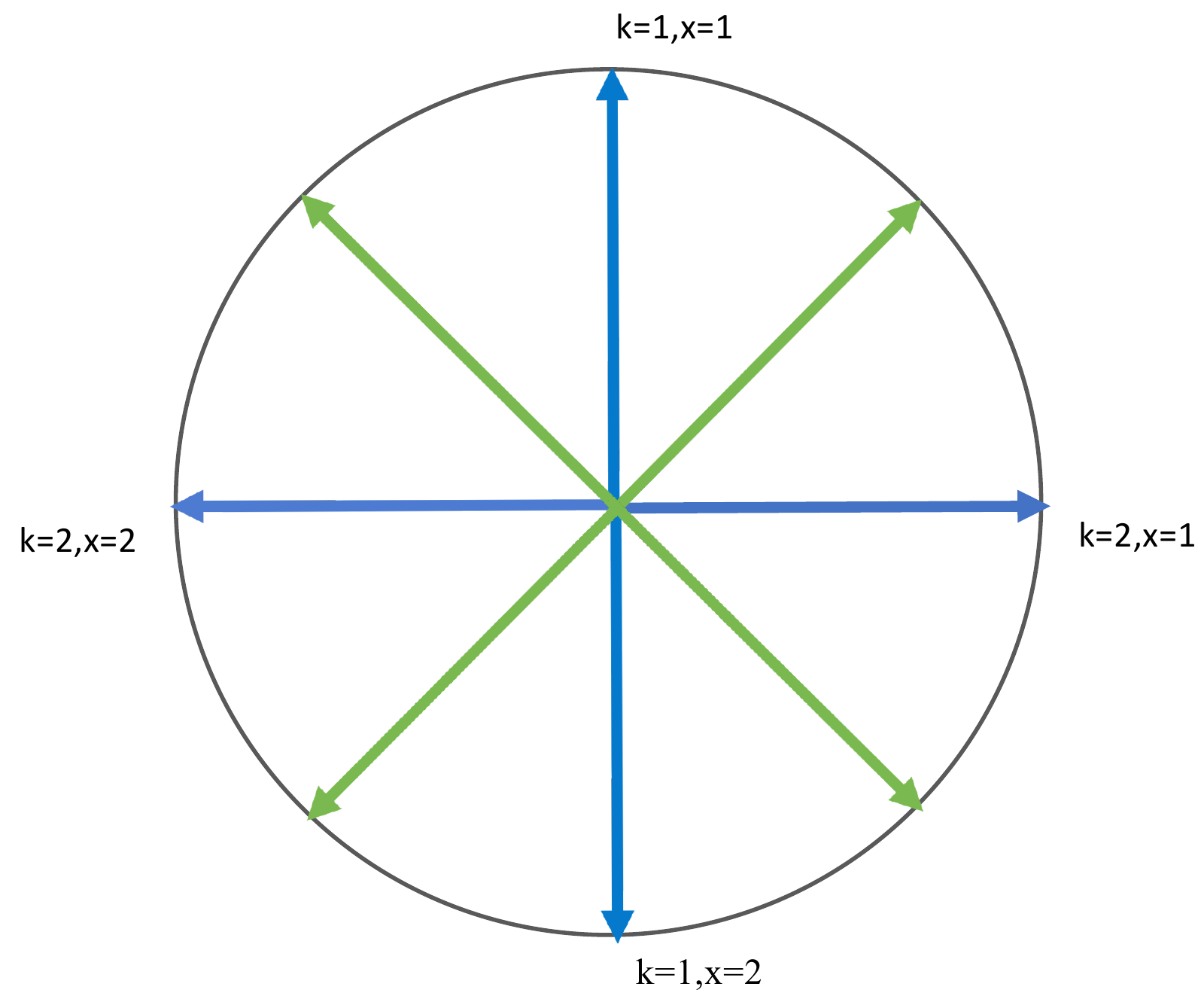}
\caption{8 vectors on xy plane of Bloch sphere, corresponding to 8 pure states defined in Eq.(\ref{four}) and Eq.(\ref{four2}). States labeled as $k=2,x=1$ and $k=2,x=2$ are eigenstates of   $\sigma_x$ and states labeled as $k=1,x=1$ and $k=1,x=2$ are eigenstates of $\sigma_y$. States labeled by green vectors are obtained by a $\pi/4$ rotation from these four states. If one can prepare these  8 states, and measure Pauli operators,  $\sigma_x, \sigma_y, \sigma_z$, one can demonstrate the lower bound $C^{\text{min}}_\text{prep}\ge \frac{2-\sqrt{2}}{8}\approx 0.07$ on inaccessible information.}
\label{Fig-guessing}
\end{center}
\end{figure}

As an example, consider states 
\be\label{frroeo}
|\psi_{(k,x)}\rangle\equiv\frac{1}{\sqrt{2}}(|0\rangle+ e^{{\rm{i}} \pi (\frac{k}{n}+x)}|1\rangle)\ , \ k=1\cdots n,\  x=1,2\ ,
\ee
whose Bloch vectors form a 2d regular polygon in the x-y equator. In particular, for $n=2$, we obtain four states
\be\label{four}
\frac{1}{\sqrt{2}}(|0\rangle+ e^{{\rm{i}} \pi (\frac{k}{2}+x)} |1\rangle)\ ,\ \  k=1,2 ; x=1,2\ ,\
\ee
whose Bloch vectors form a square in x-y plane, and $ \max_{\textbf{x}} \|\vec{n}_\textbf{x}\|=\sqrt{2}/2$. In this case  the sets of states $\{\tau_{\textbf{x},+}\}$  and  $\{\tau_{\textbf{x},-}\}$ defined in Eq.(\ref{wfqe}) coincide and are equal to 
\be\label{four2}
\frac{1}{\sqrt{2}}(|0\rangle+ e^{{\rm{i}} \pi (\frac{k}{2}+x+\frac{1}{4})} |1\rangle)\ ,\ \  k=1,2 ; x=1,2\ ,\
\ee
which can be obtained from the four  states in Eq. (\ref{four}) by applying $\pi/4$ rotation around $\hat{z}$ (See Fig.\ref{Fig-guessing}).

Therefore, assuming  preparations in $\mathcal{P} $ can prepare 4  states in Eq.(\ref{four2}), we find the non-contextuality inequality 
\be
P_\text{guess}\le Q^{\text{QM}}_\text{guess} = \frac{1+ \max_{\textbf{x}} \|\vec{n}_\textbf{x}\|}{2}=\frac{2+\sqrt{2}}{4}\approx 0.85\ .
\ee
For the optimal measurement, we have $P_\text{guess}=1$. This together with Eq.(\ref{bound99}) implies
\begin{align}\label{wfkdq}
C^\text{min}_\text{prep}&\ge \frac{P_\text{guess}-Q^{\text{QM}}_\text{guess}}{2^{n-1}}=\frac{2-\sqrt{2}}{8}\approx 0.07  \ .
\end{align}
This proves the lower bound on $C^\text{min}_\text{prep}$ in theorem \ref{Thm000}.  Note that to experimentally demonstrate this lower bound, in addition to measuring  $P_{\text{guess}}$, which   is ideally equal to one for the optimal measurement, we also need to measure a tomographically complete set of observables for 8 states, namely states in Eq.(\ref{four}) and Eq.(\ref{four2}).

Another interesting example is the case of $n=3$, where the set of states in Eq.(\ref{frroeo}) corresponds to a regular Hexagon in x-y plane. Using the symmetry of the set of vectors, it can be easily seen that  $\max_{\textbf{x}} \|\vec{n}_\textbf{x}\|=2/3$. Therefore, assuming preparations in $\mathcal{P} $ can prepare states  $\{\tau_{\textbf{x},-}\}$ or states $\{\tau_{\textbf{x},+}\}$  in Eq.(\ref{wfqe}),  we find the non-contextuality inequality 
\be
P_\text{guess}\le Q^{\text{QM}}_\text{guess}= \frac{1+ \max_{\textbf{x}} \|\vec{n}_\textbf{x}\|}{2} = \frac{5}{6} \approx 0.83\ .
\ee
Remarkably, in this case  states $\{\tau_{\textbf{x},\pm}\}$ in Eq.(\ref{wfqe}) coincide with states $\{|\psi_{(k,x)}\rangle\}$. Therefore, to demonstrate contextuality, in total we only need  $6$ different preparations. However,  it turns out that the lower bound on $C^\text{min}_\text{prep}$ in this case is weaker than the bound in Eq.(\ref{wfkdq}). In particular, using Eq.(\ref{bound99}) we find $C^\text{min}_\text{prep}\ge (P_\text{guess}-Q^{\text{QM}}_\text{guess})/{2^{n-1}}=1/24\approx 0.04$.

Interestingly, this special case of our bound has been recently found  in \cite{mazurek2016experimental}, using a completely different argument. In particular, \cite{mazurek2016experimental} shows that if in addition to PNC, a  model satisfies another condition, namely Measurement Non-Contextuality (MNC) (See Eq.(\ref{mnc})), then it predicts  $P_\text{guess}\le 5/6$. Our result shows that if one takes into account \emph{all} equivalency relations between preparations, then to derive the non-contextuality inequality $P_\text{guess}\le 5/6$, the extra assumption of Measurement Non-Contextuality is not needed  (Ref. \cite{mazurek2016experimental} claims that if MNC is not satisfied then $P_\text{guess}\le 5/6$ can be violated, even if the model  satisfies PNC. This claim is valid only if one ignores some existing equivalency relations between preparations).

\subsection{Qudit Case: Uniformly distributed states}\label{Sec:Qudit}

Next, we consider non-contextuality inequalities for a qudit with Hilbert space of dimension $D$. We assume preparations in $\mathcal{P}$ can prepare all states of the system, and measurements in $\mathcal{M}$ allow arbitrary projective measurement. 

Consider the guessing probability $P_\text{guess}$ defined in Eq.(\ref{guessing}) for the case where $d=D$ and $n\rightarrow \infty$. More precisely, suppose the set of preparations are labeled as 
\be
\mathbb{P}_{(U,j)}:\ \  U\in \text{SU}(D),\ \  j=1,\cdots,\ D\ ,
\ee
where preparation $\mathbb{P}_{(U,j)}$ prepares state
 \be
|\psi_{(U,j)}\rangle= U|j\rangle\ , 
 \ee
and $\{|j\rangle: j=1,\cdots, D\}$ is an orthonormal basis and $U$ is an arbitrary unitary. Here, unitary $U$ plays the role of the alphabet $k$ in the guessing game in Sec.\ref{Sec:Games}, and  
integer $j$ is the message that Bob should guess.  Assume unitary $U$ is  chosen uniformly at random from $\text{SU}(D)$ according to the Haar measure, and $j$ is chosen uniformly  from the set $\{1,\cdots, D\}$. 
Since for each $U$, states $\{|\psi_{(U,j)}\rangle: j=1,\cdots, D\}$ are orthogonal, it is possible to achieve $P_\text{guess}=1$  (we choose measurement $\mathbb{M}_U$  to be  the projective measurement in the basis   $\{U|j\rangle: j=1,\cdots, D\}$).

As we have seen before, PNC implies that $P_\text{guess}$ is upper bounded by  
\be
P_\text{guess}\le \frac{\alpha_\text{min}}{{D}}=Q^{\text{QM}}_\text{guess}\ ,
\ee
where $Q^{\text{QM}}_\text{guess}$, defined in Eq.(\ref{ggjf}) is the maximum probability that Bob succeeds in the modified guessing game, where he does not know the unitary $U$, but he can return his guess for the message $j$ for each possible $U$. Furthermore,   
 \be\label{wlfj}
\alpha_\text{min}\equiv\inf_{\mathbb{P}_f\in\mathcal{P}} \sup_{\textbf{j}} 2^{\mathbb{D}_\text{max}(\mathbb{P}_\textbf{j}\|\mathbb{P}_f)}=\inf_{\sigma_f} \sup_{\textbf{j}} 2^{{D}_\text{max}(\rho_\textbf{j}\|\sigma_f)}\ ,
\ee
where  $\textbf{j}: \text{U}(D)\rightarrow \{1,\cdots, D\}$  is a function from the set of unitaries acting on a $D$-dimensional space to $\{1,\cdots, D\}$, 
\be
\rho_\textbf{j}=\int dU\ U|\textbf{j}(U)\rangle\langle \textbf{j}(U)|U^\dag\ .
\ee 
and the second infimum in Eq.(\ref{wlfj}) is over the set of all density operators of the system. 

Using the qusi-convextiy of  ${D}_\text{max}$ together with the symmetry of the set of states, it can be easily shown that the infimum is achieved for $\sigma_f=I/D$, i.e., the maximally mixed state. It follows that
\bes
\begin{align}
\frac{\alpha_\text{min}}{D}&=\frac{1}{D}\inf_{\sigma_f} \sup_{\textbf{j}} 2^{{D}_\text{max}(\rho_\textbf{j}\|\sigma_f)}\\ &=\frac{1}{D} \sup_{\textbf{j}} 2^{{D}_\text{max}(\rho_\textbf{j}\|\frac{I}{D})} =\sup_{\textbf{j}} \lambda_\text{max}(\rho_\textbf{j}) \\ &=\sup_{|\eta\rangle} \int dU\  \max_{j\in\{1,\cdots, D\}}  \big|\langle\eta|U|j\rangle\big|^2\\ &=\int dU\ \max_{j\in\{1,\cdots, D\}} \big|\langle 1|U|j\rangle\big|^2\ ,
\end{align}
\ees
where $\lambda_\text{max}(\rho_\textbf{j})$ is the maximum eigenvalue of $\rho_\textbf{j}$, and we have used $\lambda_\text{max}(\rho_\textbf{j})=\sup_{|\eta\rangle} \langle\eta|\rho_\textbf{j}|\eta\rangle$ where the supremum is over all normalized pure states. The last line follows from the fact that $dU$ is the invariant measure. 

In conclusion, we find that PNC predicts
\be\label{kjbdvbd}
P_\text{guess}\le Q^{\text{QM}}_\text{guess}=\frac{\alpha_\text{min}}{D}=\int dU\ \max_{j} \big|\langle 1|U|j\rangle\big|^2\ .
\ee
Finally, we use the result of \cite{lakshminarayan2008extreme}, which shows 
\be
\int dU\ \max_{j} \big|\langle 1|U|j\rangle\big|^2= \frac{1}{D}(1+\frac{1}{2}+\cdots +\frac{1}{D})\ .
\ee

We conclude that, while according to quantum mechanics  it is possible to achieve $P_\text{guess}=1$, PNC predicts that 
\be\label{kwn}
P_\text{guess}\le  Q^{\text{QM}}_\text{guess}=\frac{1}{D}(1+\frac{1}{2}+\cdots +\frac{1}{D})\ .
\ee
Finally, it is worth mentioning that the optimal POVM which achieves $Q^{\text{QM}}_\text{guess}=\frac{1}{D}(1+\frac{1}{2}+\cdots +\frac{1}{D})$ has a simple interpretation. Recall that $Q^{\text{QM}}_\text{guess}$  defined in Eq.(\ref{ggjf}) is the maximum guessing probability in the guessing game, where Bob is given state $|\psi_{(U,j)}\rangle= U|j\rangle$, and he does not know the \emph{alphabet} $U$, but he can return a guess for the value of $j$ for each possible value of $U\in \text{SU}(D)$. In other words, he returns a function $\textbf{j}: \text{SU}(D)\rightarrow \{1,\cdots, D\}$. To achieve $Q^{\text{QM}}_\text{guess}=\frac{1}{D}(1+\frac{1}{2}+\cdots +\frac{1}{D})$, Bob can perform a projective measurement in a fixed orthonormal basis basis $\{|1\rangle,\cdots, |D\rangle\}$ and upon observing outcome $l\in \{1,\cdots, D\}$, he returns function $\textbf{j}: \textbf{j}(U)=\text{argmax} |\langle j|U|l\rangle|^2$, i.e. for alphabet $U$, he chooses $j\in \{1,\cdots, D\}$ for which   the overlap $|\langle j|U|l\rangle|^2$ is maximized.  For this strategy, the probability of correct guess is $\int dU\ \max_{j} \big|\langle 1|U|j\rangle\big|^2= \frac{1}{D}(1+\frac{1}{2}+\cdots +\frac{1}{D})=Q^{\text{QM}}_\text{guess}$.

 \section{Noise thresholds for contextuality of quantum systems}\label{Sec:noise}

 What is the maximum noise level which still allows observation of contextuality in quantum systems?  In this section, we show that using non-contextuality inequalities, such as Eq.(\ref{kwn}), we can derive lower bounds on this noise threshold. These lower bounds imply that, even in the presence of a finite amount of noise, it is still possible to demonstrate contextuality of quantum mechanics.   

To simplify the discussion, we assume noise affects  preparations but not measurements. Specifically, we assume  noise can be modeled by  a quantum channel $\mathcal{E}$, such that preparations in $\mathcal{P}$ can prepare all and only quantum states $\{\mathcal{E}(\rho)\}$ for arbitrary density operator $\rho$. On the other hand, we assume measurements in $\mathcal{M}$ include all (POVM) measurements allowed in quantum mechanics.  
 
  Clearly, in practice measurements are also imperfect and an ideal projective measurement is impossible. However, in many cases of interest, one can model the imperfections of measurements as  noise in preparations, and therefore our results are applicable.  Note that unlike theorem \ref{Thm00} and our non-contextuality inequalities, here our results rely on the validity of quantum mechanics.

 \subsection{From non-contextuality inequalities to noise thresholds}

Recall the non-contextuality inequalities $P_\text{guess}\le \min\{\frac{\alpha_\text{min}}{d}, 1-\frac{d-1}{d}\beta^{-1}_\text{min}  \} $, and the definition of the guessing probability  $P_\text{guess}$ in Eq.(\ref{guessing}). Suppose preparation $\mathbb{P}_{(k,x)}$  prepares state $\mathcal{E}(\rho_{(k,x)})$, and measurement $\mathbb{M}_k$ is described by the POVM $\{B^{(k)}_x: x=1,\cdots, d \}$. Then, the guessing probability in Eq.(\ref{guessing}) is equal to
\begin{align}\label{lne}
P_\text{guess}&=\frac{1}{n} \sum_{k=1}^n \frac{1}{d}\sum_{x=1}^d \Tr(B^{(k)}_x \mathcal{E}(\rho_{(k,x)}))\ .
\end{align}
Clearly, noise affects the guessing probability $P_\text{guess}$.  In general, as the noise becomes stronger this probability decreases. 

On the other hand, as long as  the noise channel $\mathcal{E}$ is a one-to-one map, the quantities $\alpha_\text{min}$ and $\beta_\text{min}$ remain invariant under noise, i.e.
\bes\label{lfkjpejf}
\begin{align}
 \alpha_\text{min} &\equiv  \inf_{\mathbb{P}_f\in\mathcal{\mathcal{P}}}\ \max_{\i}  2^{\mathbb{D}_\text{max}(\mathbb{P}_{\textbf{x}} \|\mathbb{P}_f)} = \inf_{\sigma_f} \max_{\textbf{x}} 2^{D_\m(\rho_{\textbf{x}}\|\sigma_f)}\  ,\\
 \beta_\text{min} &\equiv  \inf_{\mathbb{P}_f\in\mathcal{\mathcal{P}}}\ \max_{\i}  2^{\mathbb{D}_\text{max}(\mathbb{P}_f\|\mathbb{P}_{\textbf{x}}) } = \inf_{\sigma_f} \max_{\textbf{x}} 2^{D_\m(\sigma_f\|\rho_{\textbf{x}})}\ .
\end{align}
\ees
This can be seen using proposition \ref{qffq}, and is  a consequence of the fact that $\alpha_\text{min}$ and $\beta_\text{min} $ only depend on the equivalency relations between preparations, which remain invariant under a one-to-one map $\mathcal{E}$. 

Since $\alpha_\text{min},\beta_\text{min} \ge 1$, it follows that  for sufficiently strong noise, $P_\text{guess}$ will satisfy  the non-contextuality inequality $P_\text{guess}\le \min\{\frac{\alpha_\text{min}}{d}, 1-\frac{d-1}{d}\beta^{-1}_\text{min}  \} $. As we show in the following, using this approach we can obtain lower bounds on the minimum noise level  which makes the theory preparation non-contextual.

\subsection{Main result II: Noise threshold in terms of average gate fidelity}

Consider a noisy version of  the   guessing game discussed in Sec.\ref{Sec:Qudit}: Suppose  preparation $\mathbb{P}_{(U,j)}$ prepares state
 \be
\rho_{(U,j)}= \mathcal{E}(U|j\rangle\langle j|U^\dag)\ , 
 \ee
 where $\{|j\rangle: j=1,\cdots, D\}$ is an orthonormal basis for a $D$-dimensional space, and $U$ is an arbitrary unitary acting on this space. Similar to the scenario discussed  in Sec. \ref{Sec:Qudit}, assume unitary $U$  is  chosen uniformly at random from $\text{SU}(D)$ according to the Haar measure, and $j$ is chosen uniformly  from the set $\{1,\cdots, D\}$. 

In Sec.(\ref{Sec:Qudit}) we considered this guessing game in the noiseless case, i.e., when the channel $\mathcal{E}$ is the identity map, and showed  that in that case  $\alpha_\text{min}=1+\cdots +D^{-1}$.  In the previous section,  we argued that if the noise channel $\mathcal{E}$ is a one-to-one function, then  the quantities $\alpha_\text{min}$ and $\beta_\text{min}$ remain unchanged under the effect of noise. Hence, in the above scenario where  preparation  $\mathbb{P}_{(U,j)}$ prepares state $\rho_{(U,j)}$, if $\mathcal{E}$ is a one-to-one function, then $\alpha_\text{min}=1+\cdots +D^{-1}$. Therefore, the non-contextuality inequality $P_\text{guess}\le{\alpha_\text{min}}/{D}$ implies
\be\label{kjhwq}
P_\text{guess}\le \frac{\alpha_\text{min}}{D}=\frac{1}{D}(1+\cdots +\frac{1}{D})\ .
\ee
Next, we calculate $P_\text{guess}$, assuming Bob performs the projective measurement in the orthonormal basis $\{U|j\rangle: j=1,\cdots, D\}$. In this case,  the guessing probability in Eq.(\ref{lne})  is equal to 
\be
P_\text{guess}=\frac{1}{D}\int dU\ \sum_{j=1}^D \langle j|U^\dag \mathcal{E}(\rho_{(U,j)})U|j\rangle= F(\mathcal{E})\ ,
\ee
where
\be
F(\mathcal{E})\equiv \int d\eta\  \langle\eta|\mathcal{E}(|\eta\rangle\langle\eta|)|\eta\rangle
\ee
is the average gate fidelity for channel $\mathcal{E}$,  and  $d\eta$ is the uniform (Haar) measure over the set of pure states,   which satisfies the normalization  $\int d\eta=1$ \cite{horodecki1999general}.   Average gate fidelity is a standard way to quantify the noise in a quantum channel. It has a simple relation with the  entanglement fidelity \cite{horodecki1999general}, and is less than or equal to one. In particular, it is equal to one iff the channel  is the identity map, i.e., is completely noiseless.

Therefore, Eq.(\ref{kjhwq}) implies that if PNC holds then
\be
P_\text{guess}= F(\mathcal{E})\le \frac{\alpha_\text{min}}{D}=\frac{1}{D}(1+\frac{1}{2}+\cdots +\frac{1}{D})\ .
\ee
In other words, if the average gate fidelity is larger than 
\be\label{kjhjhkljhwilv}
F(\mathcal{E})> \frac{1}{D}\big(1+\frac{1}{2}+\cdots +\frac{1}{D}\big)\ ,
\ee
then it is still possible to perform a prepare-measure experiment which demonstrates violation of PNC.

As we discuss in Sec.\ref{Ex:dep}, this bound is tight in the case of the qubit depolarizing channel. 

\subsection{Necessary and sufficient condition for Preparation Non-Contextuality }\label{Sec:lkhdmain}

Consider again  the operational theory with preparations $\mathcal{P}$ which prepare all and only states $\{\mathcal{E}(\rho)\}$ for arbitrary density operator $\rho$  of a quantum system, and with measurements $\mathcal{M}$ which allow arbitrary quantum mechanical measurements. In  Appendix \ref{Sec:lkhd}, we show that this operational theory has a model satisfying PNC and convex linearity iff there exists a fixed POVM $\{E_\lambda\}$ such that for any POVM $\{B_l\}$, there exists a set $\{\zeta(l|\lambda): \zeta(l|\lambda)\ge 0, \forall\lambda: \sum_l \zeta(l|\lambda)=1\}$, and
\be\label{fwfwfff}
\Tr(\mathcal{E}(\rho) B_l)=\sum_{\lambda}  \zeta(l|\lambda)\times  \Tr(E_\lambda \rho)\ .
\ee
A quantum channel $\mathcal{E}$ which satisfies this property has a simple interpretation: Any arbitrary measurement with POVM $\{B_l\}$ on the output of the channel can be simulated by a fixed measurement $\{E_\lambda\}$,  independent of the POVM $\{B_l\}$,  on the input of the channel, followed be a stochastic map, which depends on POVM $\{B_l\}$ and generates the outcome of this measurement  based on the outcome of the fixed measurement.    \\

It is worth noting that this property also arises in the study of compatible  measurements. For any POVM $\{C_k\}_k$, consider 
the \emph{noisy} POVM $\{\mathcal{E}^\dag(C_k)\}_k$, where $\mathcal{E}^\dag$ is the adjoint of $\mathcal{E}$ defined by equation $\Tr(Y\mathcal{E}(X))=\Tr(\mathcal{E}^\dag(Y) X)$ for arbitrary pair of operators $X$ and $Y$.  If the noise channel satisfies the above property, then for any sets of POVM's $\{C^{(r)}_k\}_k: r=1,\cdots, R$ , their noisy versions, i.e.,  $\{\mathcal{E}^\dag(C^{(r)}_k)\}_k: r=1,\cdots, R$,  can be measured simultaneously: one first performs the fixed POVM $\{E_\lambda\}_\lambda$ in Eq.(\ref{fwfwfff}). Then, to simulate each measurement $\{\mathcal{E}^\dag(C^{(r)}_k)\}_k$, one applies a proper stochastic map $\zeta^{(r)}$ to the outcome of the fixed measurement.  The connection between measurement compatibility and non-contextuality has been previously discussed in \cite{liang2011specker}.

\subsection{Measurement and Preparation Non-Contextuality hold iff the noise channel is entanglement-breaking }\label{Sec:lkhd2main}

So far, in this paper we have only focused on the contextuality of preparations. To understand the effect of noise on quantum systems, it is also interesting to consider the notion of non-contextuality for measurements, as  defined in \cite{spekkens2005contextuality}: Suppose for all preparations in $\mathcal{P}$ the probability of outcome $m$ of measurement $\mathbb{M}$ is equal to the   probability of outcome $m'$ of measurement $\mathbb{M}'$, i.e. $\forall  \mathbb{P}\in\mathcal{P}:\  P(m| \mathbb{P}, \mathbb{M})=P(m'| \mathbb{P}, \mathbb{M}')\ .$ Measurement Non-Contextuality (MNC) states that, for such measurement outcomes,  the corresponding response functions should be identical, i.e.  $\forall\lambda\in\Lambda: \xi_{\mathbb{M}}(m|\lambda)=\xi_{\mathbb{M}'}(m'|\lambda)$. 

More generally, consider the equivalency relation 
\be
\forall  \mathbb{P}\in\mathcal{P}:\  \sum_j p_j\ P(m_j | \mathbb{P}, \mathbb{M}_j)=\sum_j p'_j\ P(m'_j| \mathbb{P}, \mathbb{M}_j')\ ,
\ee  
where $\{p_i\}$ and  $\{p'_j\}$ are arbitrary probability distributions. Then, MNC implies
\be\label{mnc}
\forall  \mathbb{P}\in\mathcal{P}:\  \sum_i p_i\ \xi_{\mathbb{M}_i}(m_i|\lambda)=\sum_j p'_j\ \xi_{\mathbb{M}'_j}(m'_j|\lambda)\ .
\ee  
This condition is the counterpart of Eq.(\ref{elnk}) for preparations. If an operational theory does not have  a model satisfying this condition,  we say the theory is measurement contextual.    \\

In  Appendix \ref{Sec:lkhd2},  we show that the operational theory whose preparations prepare all and only states $\{\mathcal{E}(\rho)\}$ for arbitrary density operators $\rho$  of a quantum system  and whose measurements $\mathcal{M}$ allow arbitrary measurements has a model satisfying both PNC and MNC iff the channel $\mathcal{E}$ is entanglement-breaking.

\subsection{Example: Depolarizing channel}\label{Ex:dep}

Consider the special case where the noise is described by a depolarizing channel
\be
\mathcal{E}(\rho)= (1-p) \rho+ p \frac{I}{D}\ ,
\ee
where $0\le p\le 1$, and $\frac{I}{D}$ is the maximally mixed state of  a $D$-dimensional system.

It can be easily seen that the average gate fidelity of this channel is $1- p(D-1)/D$. Furthermore, this channel is entanglement-breaking iff $p\ge D/(D+1)$  \cite{horodecki1999reduction, wilde2013quantum}. Therefore, using Eq.(\ref{kjhjhkljhwilv}) we find  
\bes
\begin{align}
&p< \frac{D-[1+\cdots+D^{-1}]}{D-1} &&\Longrightarrow\  \text{Not PNC}\label{ksks1}\\
&p\ge \frac{D}{D+1} &&\Longleftrightarrow\   \text{Both PNC and MNC}\ \label{ksks2}.
\end{align}
\ees
In the limit of large $D$, this means that for $p<1-\log D/D$, the theory is preparation contextual and for  $p\ge 1-1/D$ is preparation and measurement non-contextual. 

In Appendix \ref{Sec:ex-dep}, we show that the bound in Eq.(\ref{ksks1}) is tight
in the case of a single qubit ($D=2$), i.e., for $p \ge 1/2$ there exist models satisfying PNC (as well as convex-linearity).  Note that according to Eq.(\ref{ksks2}) such models cannot satisfy MNC, unless $p\ge 2/3$ (because for $p<2/3$ the depolarizing channel is not Entanglement-Breaking).

\section{Discussion}\label{Sec:Discus}
We introduced the notions of inaccessible information of a model,  $C_\text{prep}$, and  inaccessible information of an operational theory,  $C^\text{min}_\text{prep}=\inf_\text{Models} C_\text{prep}$.  Choosing the model with the lowest $C_\text{prep}$ can be thought of as an information theoretic model selection criterion, which prefers  models with higher \emph{efficiency}, and imposes preparation non-contextuality if possible. In a sense this can be thought of as a relaxed version of the Leibniz's principle of identity of indiscernibles \cite{spekkens2019ontological}. For any operational theory the value of $C^\text{min}_\text{prep}$ quantifies a certain notion of non-classicality associated to preparation contextuality.    

We found a method  to experimentally demonstrate a lower bound on $C^\text{min}_\text{prep}$.  Any such lower bound on $C^\text{min}_\text{prep}$ can also be interpreted as a violation of a non-contextuality inequality. 
 In the example discussed in Fig.\ref{Fig-guessing}, by  preparing 8 different pure states, and measuring the Pauli operators, ideally one can demonstrate the lower bound $C^\text{min}_\text{prep}\ge 0.07$ for the operational theory corresponding to a single qubit (In fact, as we discussed in Sec.\ref{Sec:qubit}, preparing 6 pure states suffices  to demonstrate $C^\text{min}_\text{prep}\ge 0.04$).   
 
We also introduced the notions of operational total variation distance and operational max-relative entropy, which could be of independent interest. Note that although we introduced these concepts in the context of operational theories,  they can also be defined in the framework of Generalized Probabilistic Theories (GPT) \cite{hardy2001quantum, barrett2007information}.  For any operational theory, one can define a  GPT by removing  the redundancies due to equivalency between different preparations and between different measurements. In particular, each equivalency class of preparations defines a \emph{state}  in the corresponding GPT. Therefore, since the operational total variation distance  $\mathbb{d}_\text{prep}(\mathbb{P}_a, \mathbb{P}_b)$ and the operational max-relative entropy $\mathbb{D}_\text{max}(\mathbb{P}_a, \mathbb{P}_b)$ only depend on the equivalency classes of preparations $\mathbb{P}_a$ and $\mathbb{P}_b$, they can also be thought of as functions of states $\omega_a$ and  $\omega_b$ associated to these preparations, i.e.
\begin{align}
\tilde{D}_\text{max}(\omega_a, \omega_b)\equiv \mathbb{D}_\text{max}(\mathbb{P}_a, \mathbb{P}_b)\ ,\\
\tilde{d}_\text{prep}(\omega_a, \omega_b)\equiv \mathbb{d}_\text{prep}(\mathbb{P}_a, \mathbb{P}_b)\ ,
\end{align}
where $\tilde{D}_\text{max}$ and $\tilde{d}_\text{prep}$ are measures of distinguishability of states in the GPT. This implies
\be
\tilde{D}_\text{max}(\omega_a, \omega_b)=\tilde{d}_\text{prep}(\omega_a, \omega_b)=0 \ \Longleftrightarrow \ \omega_a=\omega_b\ .
\ee

 Many interesting questions are left open in this work. For instance, we found that for any quantum system with a finite-dimensional Hilbert space $C^\text{min}_\text{prep}$ is strictly less than one, and for a single qubit $C^\text{min}_\text{prep}\le 1/2$. But, the actual value of  $C^\text{min}_\text{prep}$ as a function of dimension remains unknown.  Furthermore, the lower bounds on $C^\text{min}_\text{prep}$ in Eq.(\ref{aefk})  do not seem to be tight. 
Also, given the close relation between contextuality and negativity  \cite{spekkens2008negativity}, it is  interesting to understand the connection between inaccessible information  $C^\text{min}_\text{prep}$ and measures of negativity.  In future work, we will study the inaccessible information  $C^\text{min}_\text{prep}$ in the context of parity-oblivious multiplexing \cite{spekkens2009preparation}, which has been shown to be closely related to preparation contextuality.

\section*{Acknowledgments}

I  am grateful to Shiv Akshar Yadavalli and Robert Spekkens for reading the manuscript carefully and providing many useful comments. This work was supported by NSF grant FET-1910859.

\bibliography{Ref_2018}

\onecolumngrid
\appendix

\newpage

\maketitle
\vspace{-5in}
\begin{center}

\Large{Supplementary Material }
\end{center}
\appendix



\section{A universal ontological model}\label{App:A}

Here, we present a universal ontological model, which can be constructed for any operational theory. This model is sometimes called the \emph{Kitchen-sink}   model \cite{ruebeck2018epistemic}.  The ontic states of the model, denoted by $\Lambda=\{\lambda\}$, are the list of all possible outcomes of all measurements, i.e. $\lambda=(\lambda_1, \lambda_2,\cdots )$, where  $\lambda_k$ is an outcome of measurement $\mathbb{M}_k$, and $\mathcal{M}=\{\mathbb{M}_1, \mathbb{M}_2, \cdots\}$ is the set of all measurements  in the operational theory (For example, if $\mathcal{M}$ consists of $n$ two-outcome measurements, there will be $2^n$ ontic states). The probability distribution associated to  preparation $\mathbb{P}$ and the response function associated to outcome $m$ of measurement $\mathbb{M}$ are, respectively,  defined by 
\bes\label{assign}
\begin{align}
\mu_{\mathbb{P}}(\lambda) &=\prod_{k=1}^n P(\lambda_k| \mathbb{M}_k, \mathbb{P})\ ,\\
\xi_{\mathbb{M}_k}(m|\lambda) &=\delta_{m,\lambda_k}\ ,
\end{align}
\ees
where $\lambda_k$ is the $k$'th element of $\lambda$. It can be easily seen that this model reproduces  the statistics of the operational theory via Eq.(\ref{def1}). However,  it does not satisfy the convex-linearity criterion. This is a consequence of the fact that $\mu_{\mathbb{P}}(\lambda)$ is a non-linear function of probabilities $P(m_k| \mathbb{M}_k, \mathbb{P})$.

However, a modified version of this model \emph{does} satisfy the convex-linearity criterion: Suppose we only define the assignments in Eq.(\ref{assign}) in the case of extremal (pure) preparations and measurements  and then use convex-linearity to extend them to all preparations and measurements. Then, the model will be convex-linear  by construction.

The above recipe for constructing an ontological model    
 is universal, i.e., can be applied to any operational theory, including quantum mechanics. In fact, in some cases this is the most efficient model for describing an operational theory. This is the case, for instance, if each measurement $\mathbb{M}_k$ is performed on a separate system $k$, and if different   systems can be prepared independently.  However, in general, this model is not \emph{economical}, i.e. the ontological description of preparations and measurements contain extra information which are operationally irrelevant.

\section{A new proof of preparation contextuality of quantum mechanics}\label{App:B}

Here, we present a simple argument which proves an ideal quantum mechanical system is preparation contextual, i.e.,  cannot be described by an ontological model satisfying PNC.  We present the proof in the case of a qubit. The argument relies on the assumption that a qubit can have 2 distinct pairs of orthogonal states (In general, if the Hilbert space is not 2-dimensional, one can consider states restricted to a 2-dimensional subspace).  

Let $\Lambda=\{\lambda\}$ be the set of ontic states associated to a qubit. Without loss of generality, we  assume all states in $\Lambda$  have a non-zero probability for some states of the qubit. More precisely, we assume $\Lambda$ is the set of all ontic states $\lambda$, where each $\lambda$ belongs to the support of the probability distribution associated to a quantum state (If an ontic state has probability zero for all states then it is irrelevant and we can remove it from the model).

  Any mixed qubit state $\rho$ is a full-rank density operator. Therefore, given any other density operator $\sigma$, a mixed density operator  $\rho$ can be written as the convex combination $\rho=p \sigma+(1-p) \sigma'$, where  $\sigma'$ is another  density operator and $0<p< 1$, i.e., $p$ is strictly larger than zero.
 Therefore, to prepare $\rho$, we can prepare $\sigma$ with probability $p$ and $\sigma'$ with probability $1-p$.  By convex-linearity, the probability distribution associated to such preparation  is the convex combination of the distributions associated to $\sigma$ and $\sigma'$, with weights $p$ and $1-p$, respectively. 
PNC implies that this is also the distribution associated to $\rho$. Therefore, if the ontic state $\lambda$ is in the support of the probability distribution associated to $\sigma$, then it should also be in the support of the probability distribution associated to $\rho=p \sigma+(1-p) \sigma'$. But, since $\sigma$ is arbitrary and $p>0$ this implies that for any mixed state $\rho$ the corresponding  probability distribution  should have  full support, i.e., its support should be equal to $\Lambda$.

Consider an arbitrary ontic state $\lambda\in \Lambda$. Suppose there are two distinct  pure states which both assign probability zero to $\lambda$. Then, their mixture is a full-rank qubit density operator which  assigns probability zero to $\lambda$, in contradiction with the above result. Therefore, we conclude that for any given ontic state $\lambda\in \Lambda$ there is, at most, one pure state with no support on $\lambda$. Let us denote this state by $\psi_\lambda$ and the pure state orthogonal to this state by $\psi_\lambda^\perp$.  Any other pair of states $\phi_1$ and $\phi_2$ will assign a non-zero probability to $\lambda$. Assuming the set of ontic states are finite, this means that they both associate a finite (larger than zero) probability to the ontic state $\lambda$, and therefore their corresponding probability distributions have a non-zero overlap.   This immediately implies that the arbitrary pair $\phi_1$ and $\phi_2$  cannot be perfectly distinguishable. However, according to quantum mechanics, if $\phi_1$ and $\phi_2$ are orthogonal then they are perfectly distinguishable. 

In conclusion, we find that:  Assuming the set of ontic states $\Lambda=\{\lambda\}$ is finite, then preparation non-contextuality implies that a qubit can have, at most, one pair of perfectly distinguishable states, in contradiction with quantum mechanics.

Although this argument provides a simple proof of contextuality of quantum mechanics,  it is not a  quantitive statement, i.e.,  it does not determine how strongly  non-contextuality is violated in quantum mechanics. Furthermore, it is not experimentally testable, because in a real  experiment one can never prepare a pure state or perform a perfect projective measurement. Also, it is not clear how the argument can be extended to the case of infinite ontic states. The result presented in theorem \ref{Thm00}, can be thought of as a more sophisticated version of this argument, which overcomes all the aforementioned shortcomings.  

\section{Upper bound on the inaccessible information  (Proof of Eq.(\ref{kkh}))}\label{App:2020}
As we discussed below theorem \ref{Thm000}, for certain ontological models,  to determine the inaccessible information, we can restrict our attention to the case of ensembles of pure states, i.e.
\be\label{App:lejfpejf}
C_\text{prep}\equiv   \sup d_{\text{TV}}\Big(\sum_i p_i \mu_i\ ,\ \sum_j p'_j \mu'_j \Big)\ ,
\ee
where $\mu_i$ and $\mu'_j$  are the probability distributions associated to 
pure states $\psi_i$ and $\psi'_j$, and the supremum is over all pairs of ensembles $\{p_i, \psi_i\}$ and $\{p'_j, \psi'_j\}$, which satisfy 
\be\label{App:oijw2}
\sum_i p_i |\psi_i\rangle\langle\psi_i|=\sum_j p'_j |\psi'_j\rangle\langle\psi'_j|\ .
\ee
In the following, we derive an upper bound on this quantity.

First, note that for any $\delta \ge 0$, we have
\bes\label{App:oijw}
\begin{align}
d_{\text{TV}}\Big(\sum_i p_i \mu_i,\sum_j p'_j \mu'_j \Big)&\le \sum_{i,j} p_i p'_j \ d_{\text{TV}}(\mu_i,\mu'_j )\\ &\le 1-\sum_{i,j: |\langle\psi'_j|\psi_i\rangle|^2\ge \delta} p_i p_j'\  [1-d_{\text{TV}}(\mu_i,\mu'_j )] \\ &\le 1- \min_{|\langle\psi'|\psi\rangle|^2\ge \delta}[1-d_{\text{TV}}(\mu,\mu')]\times \sum_{i,j: |\langle\psi'_j|\psi_i\rangle|^2\ge \delta} p_i p_j'\ ,
\end{align}
\ees
where the summations in the second and third lines are over all $i,j$ for which $|\langle\psi'_j|\psi_i\rangle|^2\ge \delta$. Here, the 
first inequality follows from the triangle inequality and the second inequality follows using $d_{\text{TV}}(\mu_i,\mu'_j )\le 1$.

Next, recall that for any density operator $\rho$ on a $D$-dimensional Hilbert space, $\Tr(\rho^2)\ge D^{-1}$. Combining this with  Eq.(\ref{App:oijw2}), we find 
$\sum_{i,j} p_i p_j' \  |\langle\psi'_j|\psi_i\rangle|^2 \ge {D}^{-1}$. 
Since $0\le |\langle\psi'_j|\psi_i\rangle|^2\le 1$, this  immediately implies that for any $\delta$ in the interval $0\le\delta\le D^{-1}$, 
\be\label{owhr}
\sum_{i,j: |\langle\psi'_j|\psi_i\rangle|^2\ge \delta} p_i p_j' \ \ge \frac{1}{D}-\delta\ .
\ee
Combining this bound with Eq.(\ref{App:oijw}), we find 
\be
C_{\text{prep}} \le 1- \big[D^{-1}-\delta\big]\times\big[1-\max_{|\langle\psi'|\psi\rangle|^2\ge \delta} d_{\text{TV}}(\mu,\mu')\big] \   ,
\ee
which reduces to Eq.(\ref{kkh}) in the case of $\delta=(2D)^{-1}$.

\section{Proof of lemma \ref{lem}}\label{App:lemma}

Recall the statement of lemma \ref{lem}: Let $\mu_z$ be the probability associated to preparation $\mathbb{P}_z\in \mathcal{P}$, where $z=1,\cdots , N$. Then, 
\bes
\begin{align}
\sum_{\lambda} \max_z \mu_z(\lambda) &\le {\gamma_1} (1 + N\times C_\text{prep})\ \label{lejf02}, \\
\sum_\lambda \min_{z}  \mu_{z}(\lambda)&\ge \gamma_2^{-1}-  N\times C_\text{prep}\  \label{lejf03},
\end{align}
\ees
where $C_\text{prep}\equiv \sup_{ \mathbb{P}_a \sim \mathbb{P}_b} d_\text{TV}(\mu_a,\mu_b)$ is the  inaccessible information of the  model, and  
\begin{align}
\gamma_1 \equiv \inf_{\mathbb{P}_f\in\mathcal{\mathcal{P}}}\ \max_{z}  2^{\mathbb{D}_\text{max}(\mathbb{P}_z \|\mathbb{P}_f)}\ ,\\
\gamma_2 \equiv \inf_{\mathbb{P}_f\in\mathcal{\mathcal{P}}}\ \max_{z}  2^{\mathbb{D}_\text{max}(\mathbb{P}_f \|\mathbb{P}_z)}\ .
\end{align}

We start by proving Eq.(\ref{lejf02}). For any $\mathbb{P}_f\in\mathcal{P}$, let  $w_\ast\in [0,1]$ be
\be
 w_\ast=\min_{z\in\{1,\cdots, N\}}  2^{-\mathbb{D}_\text{max}(\mathbb{P}_\emph{z} \|\mathbb{P}_f )}\ .
\ee
Using the definition of $\mathbb{D}_\text{max}$ this means that for any $w\in [0, w^\ast)$ the following holds: For each  $z\in \{1,\cdots , N\}$ there exists a preparation $\mathbb{P}'_z$, which satisfies 
\begin{align} \label{constraints2}
\Big\{ (w, \mathbb{P}_{z}) , ([1-w] , \mathbb{P}'_z)\Big\}\sim \mathbb{P}_f\ ,
\end{align}
i.e., the preparation in which we apply  $ \mathbb{P}_z$ with probability $w$ and  $ \mathbb{P}'_z$ with probability $(1-w)$ is equivalent to  preparation $\mathbb{P}_f$. 

Using the convex-linearity assumption, the first preparation is described by $w \mu_z+(1-w)\mu'_z $, where  $\mu_z$ and $\mu'_z $ are, respectively, the distributions associated to $\mathbb{P}_z$ and $ \mathbb{P}'_z$.  Suppose the distribution associated to 
 $ \mathbb{P}_f$ is $ \mu_f$. If the ontological model satisfies PNC then these two distributions coincide. In general, however, these distributions could be different, but their total variation distance is bounded by 
\be
d_\text{TV}(w \mu_z+(1-w)\mu'_z, \mu_f)\le {C}_\text{prep} \ ,
\ee
where ${C}_\text{prep}$ is the inaccessible information  of the ontological model. This means that
\be\label{ed}
\mu_f(\lambda)=[w \mu_z(\lambda) +(1-w)\mu'_{z}(\lambda)]+\delta_z(\lambda)\ ,
\ee
where 
\be
\frac{1}{2}\sum_\lambda |\delta_z(\lambda)|\le  {C}_\text{prep}\ ,
\ee
and 
\be
\sum_\lambda \delta_z(\lambda)=0\ .
\ee
This implies that for any $\Lambda'\subseteq \Lambda$,
\be\label{kw}
\Big|\sum_{\lambda\in \Lambda'} \delta_z(\lambda)\Big|\le {C}_\text{prep} \ .
\ee
Now suppose we partition $\Lambda$ to $N$  disjoint subsets $\Lambda_y\subset \Lambda$, corresponding to $y\in\{1,\cdots,N\}$, such that $\Lambda=\cup_y \Lambda_y$ and 
 \be
 \forall \lambda\in\Lambda_y, \forall z\in\{1,\cdots,N\}:\ \   \mu_{y}(\lambda)= \max_{z}\mu_{z}(\lambda)\ ,
 \ee
i.e. for each $\lambda\in  \Lambda_y$, the maximum of $\mu_{z}(\lambda)$, as a function of $z$  is achieved for $z=y$ (If there are multiple   $z\in\{1,\cdots, N\}$  which maximize $\mu_{z}(\lambda)$, we  pick one of them, e.g., the smallest $z$). Then,  
\bes\label{klhjef}
\begin{align}
\sum_{\lambda\in\Lambda} \max_{z}  \mu_{z}(\lambda)&=\sum_{y=1}^n \sum_{\lambda\in\Lambda_y}    \max_{z}  \mu_{z}(\lambda)\\ &=  \sum_{y=1}^n \sum_{\lambda\in \Lambda_y} \mu_{y}(\lambda)\\ &=  \frac{1}{w}\sum_y \sum_{\lambda\in \Lambda_y}\big[\mu_f(\lambda)-(1-w)\mu'_y(\lambda)- \delta_y(\lambda)\big]\\ &\le  \frac{1}{w} \sum_y \sum_{\lambda\in \Lambda_y}  [\mu_f(\lambda) -\delta_y(\lambda)] \\ &\le \frac{1}{w}[1+ \sum_y \Big|\sum_{\lambda\in \Lambda_y} \delta_y(\lambda)\Big|] \\ &\le \frac{1+N \times  C_\text{prep}}{w}\ ,
\end{align}
\ees
where to get the first equality we have used the fact that $\Lambda$ is partitioned to $\cup_y \Lambda_y$, to get the second line we have used the fact that for $\lambda\in\Lambda_y$, $\max_{z}  \mu_{z}(\lambda)=\mu_{y}(\lambda)$, to get the third line we have used Eq.(\ref{ed}), to get the fourth line we have used $\mu'_y(\lambda)\ge 0$, to get the fifth line we have again used the fact that $ \Lambda_y$ are disjoints and   $\Lambda=\cup_y \Lambda_y$,  and $\sum_\lambda \mu_f(\lambda) =1$, and finally to get the last line we have used Eq.(\ref{kw}) together with the fact that there are $N$ different regions  $\Lambda_y$ corresponding to $y\in\{1,\cdots,N\}$.

Next, recall that the above bound holds for any  $\mathbb{P}_f\in\mathcal{P}$ and  $w<w_\ast=\min_{z}  2^{-\mathbb{D}_\text{max}(\mathbb{P}_{z} \|\mathbb{P}_f )}$. Therefore, taking the infimum of the bound in Eq.(\ref{klhjef}) over all possible values of $w<w_\ast$, we find
\bes
\begin{align}
\sum_\lambda \max_{z}  \mu_{z}(\lambda)&\le \frac{1+N\times  C_\text{prep}}{w_\ast} \\ &=   \frac{1+N\times  C_\text{prep}}{ \min_{z}  2^{-\mathbb{D}_\text{max}(\mathbb{P}_{z} \|\mathbb{P}_f )}} \\ &
=\max_{z} 2^{\mathbb{D}_\text{max}(\mathbb{P}_{z} \|\mathbb{P}_f )} \times [1+ N\times  C_\text{prep}]\ .
\end{align}
\ees
Next, note that this holds for all $\mathbb{P}_f\in\mathcal{P}$. Therefore, taking the infimum of the right-hand side over all preparations $\mathbb{P}_f$, we find
\be\label{ef61}
\sum_\lambda \max_{z}  \mu_{z}(\lambda)\le \gamma_1\times [1+N\times  C_\text{prep}]\ .
\ee
This proves Eq.(\ref{lejf02}). 

Next, we prove Eq.(\ref{lejf03}) using a similar argument. For any $\mathbb{P}_f\in\mathcal{P}$, let  $v_\ast\in [0,1]$ be
\be
 v_\ast\equiv \min_{z}  2^{-\mathbb{D}_\text{max}(\mathbb{P}_f \|\mathbb{P}_{z} )}\ .
\ee
Using the definition of $\mathbb{D}_\text{max}$, for any $v\in [0, v^\ast)$ the following holds: For each $z\in\{1,\cdots, N\}$, there exists a preparation $\mathbb{P}'_z$, such that
\begin{align}
\Big\{ (v, \mathbb{P}_{f}) , ([1-v] , \mathbb{P}'_z)\Big\}\sim \mathbb{P}_z\ ,
\end{align}

Again, following the same argument we used before, we find
\be
d_\text{TV}(v \mu_f+(1-v)\mu'_z\ ,\ \mu_z)\le {C}_\text{prep} \ ,
\ee
which implies
\be
\mu_z(\lambda)=[v \mu_f(\lambda)+(1-v) \mu'_{z}(\lambda)]+\delta_z(\lambda)\ ,
\ee
where $\frac{1}{2}\sum_\lambda |\delta_z(\lambda)|\le  {C}_\text{prep}$, and 
$\sum_\lambda \delta_z(\lambda)=0\ $. Therefore, for any $\Lambda'\subseteq \Lambda$, 
\be
\Big|\sum_{\lambda\in\Lambda'} \delta_z(\lambda)\Big|\le {C}_\text{prep}\ .
\ee 
Then, we use the technique  used above in Eq.(\ref{klhjef}), i.e., we partition $\Lambda$ to $N$ subsets  $\Lambda_y: y\in \{1,\cdots, N\}$, such that
 \be
 \lambda\in \Lambda_y: \min_z  \mu_{z}(\lambda)=\mu_{y}(\lambda)\ .
 \ee 
Using this technique, we find 
\bes
\begin{align}
\sum_\lambda \min_{z}  \mu_{z}(\lambda)&= \sum_{y} \sum_{\lambda\in \Lambda_y} \mu_y(\lambda)\\ &=  \sum_y  \sum_{\lambda\in \Lambda_\textbf{y}}\big[v \mu_f(\lambda)+(1-v)\mu'_{y}(\lambda)+ \delta_{y}(\lambda)\big]\\ &\ge   \sum_{{y}} \sum_{\lambda\in \Lambda_{y}}  v \mu_f(\lambda) +\delta_{y}(\lambda) \\ &\ge v- \sum_{{y}} \Big|\sum_{\lambda\in \Lambda_{y}} \delta_{y}(\lambda)\Big| \\ &\ge v-N\times  C_\text{prep}\ .
\end{align}
\ees
Next, recall that the above holds for any $v<v_\ast= \min_{z}   2^{-\mathbb{D}_\text{max}(\mathbb{P}_f\|\mathbb{P}_{z})} $. Therefore, 
\be
\sum_\lambda \min_{z}  \mu_{z}(\lambda)\ge v_\ast-N\times  C_\text{prep}\ .
\ee
Next, note that this holds for all $\mathbb{P}_f\in\mathcal{P}$. Therefore, taking the supremum  over all preparations $\mathbb{P}_f$,  we find
\begin{align}
\sum_\lambda \min_{z}  \mu_{z}(\lambda)&\ge \sup_{\mathbb{P}_f\in\mathcal{P}}\min_{z}   2^{-\mathbb{D}_\text{max}(\mathbb{P}_f\|\mathbb{P}_{z})} -N \times  C_\text{prep}\ \\  &=\frac{1}{\inf_{\mathbb{P}_f\in\mathcal{P}}\max_{z}   2^{\mathbb{D}_\text{max}(\mathbb{P}_f\|\mathbb{P}_{z})}} -N \times  C_\text{prep}\\  &=\gamma_2^{-1}-N \times  C_\text{prep}\ .
\end{align}
This proves Eq.(\ref{lejf03}) and completes the proof of lemma  \ref{lem}.

\section{Operational total variation distance}\label{App:D}

\subsection{Proof of triangle inequality for operational total variation distance }

Recall the definition of the operational total variation distance, 
\begin{align}
&\mathbb{d}_{\text{prep}}(\mathbb{P}_a, \mathbb{P}_b)\equiv \inf_{q\ge 0} 
\frac{q}{1-q}: \exists \mathbb{P}_{a'},\mathbb{P}_{b'}\in\mathcal{P},   \Big\{(1-q, \mathbb{P}_a), (q , \mathbb{P}_{a'})\Big\} \sim \Big\{(1-q, \mathbb{P}_b), (q, \mathbb{P}_{b'}) \Big\}\ \ ,
\end{align}
which is equivalent to
\begin{align}\label{second:App}
&\mathbb{d}_{\text{prep}}(\mathbb{P}_a, \mathbb{P}_b)\equiv \inf  
\Big\{r\ge 0: \exists\ \mathbb{P}_{a'},\mathbb{P}_{b'}\in\mathcal{P},  \forall \mathbb{M}, m:\   P(m|\mathbb{M}, \mathbb{P}_a)-P(m|\mathbb{M}, \mathbb{P}_{b})= r  \big[P(m|\mathbb{M}, \mathbb{P}_{b'})-P(m|\mathbb{M}, \mathbb{P}_{a'})\big] \Big\}\  . 
\end{align}
To prove the triangle inequality, we use the definition in Eq.(\ref{second:App}). Suppose  there exists preparations $\mathbb{P}_{a'}$ and $\mathbb{P}_{b'}$ such that
    \begin{align}\label{76lwfj}
 \forall \mathbb{M}, m:\  \   &P(m|\mathbb{M}, \mathbb{P}_a)-P(m|\mathbb{M}, \mathbb{P}_{b})= r_{ab} \big[  P(m|\mathbb{M}, \mathbb{P}_{b'})-P(m|\mathbb{M}, \mathbb{P}_{a'})\big]\ .
 \end{align}
Similarly, suppose there exists preparations $\mathbb{P}_{b''}$ and $\mathbb{P}_{c'}$ such that
    \begin{align}\label{76lwfj2}
 \forall \mathbb{M}, m:\  \      &P(m|\mathbb{M}, \mathbb{P}_b)-P(m|\mathbb{M}, \mathbb{P}_{c})=r_{bc} \big[  P(m|\mathbb{M}, \mathbb{P}_{c'})-P(m|\mathbb{M}, \mathbb{P}_{b''})\big]\ .
 \end{align}
 Let 
 \be
 \mathbb{P}_{c''}\equiv \Big\{(\frac{r_{ab}}{r_{ab}+r_{bc}}, \mathbb{P}_{b'}), (\frac{r_{bc}}{r_{ab}+r_{bc}}, \mathbb{P}_{c'}) \Big\}\ ,
 \ee
i.e., $ \mathbb{P}_{c''}$ is  the preparation in which with probability $\frac{r_{ab}}{r_{ab}+r_{bc}}$ we apply $\mathbb{P}_{b'}$ and with probability  $\frac{r_{bc}}{r_{ab}+r_{bc}}$ we apply  $\mathbb{P}_{c'}$.   
 Similarly, let
  \be
 \mathbb{P}_{a''}\equiv \Big\{(\frac{r_{ab}}{r_{ab}+r_{bc}}, \mathbb{P}_{a'}), (\frac{r_{bc}}{r_{ab}+r_{bc}}, \mathbb{P}_{b''}) \Big\}\ ,
 \ee
 i.e., $\mathbb{P}_{a''}$ is the preparation in which with probability $\frac{r_{ab}}{r_{ab}+r_{bc}}$ we apply $\mathbb{P}_{a'}$ and with probability  $\frac{r_{bc}}{r_{ab}+r_{bc}}$ we apply  $\mathbb{P}_{b''}$. Then, by adding Eq.(\ref{76lwfj}) and Eq.(\ref{76lwfj2})  we find
   \begin{align}
 \forall \mathbb{M}, m:\  \     P(m|\mathbb{M}, \mathbb{P}_a)-P(m|\mathbb{M}, \mathbb{P}_{c})= [r_{ab}+r_{bc}]\   \big[  P(m|\mathbb{M}, \mathbb{P}_{c''})-P(m|\mathbb{M}, \mathbb{P}_{a''})\big]\ .
 \end{align}
Comparing this with Eq.(\ref{second:App}), we find that 
\be
r_{ab}+r_{bc} \ge \mathbb{d}_{\text{prep}}(\mathbb{P}_a, \mathbb{P}_c)\ .
\ee
This holds for any  $r_{ab}$ and $r_{bc} $ which satisfy  Eq.(\ref{76lwfj}) and 
 Eq.(\ref{76lwfj2}), respectively, for some preparations $\mathbb{P}_{a'}, \mathbb{P}_{b'}, \mathbb{P}_{c'}$, and $ \mathbb{P}_{b''}$. Taking the infimum over $r_{ab}$ and $r_{bc} $  with this property, we find
 \be
 \mathbb{d}_{\text{prep}}(\mathbb{P}_a, \mathbb{P}_b)+\mathbb{d}_{\text{prep}}(\mathbb{P}_b, \mathbb{P}_c) \ge \mathbb{d}_{\text{prep}}(\mathbb{P}_a, \mathbb{P}_c) ,
 \ee
  which is the triangle inequality.

\subsection{Proof of Proposition \ref{Thm0}}

Recall the statement of proposition \ref{Thm0}: 
Let   $\rho_a$ and $\rho_b$ be the density operators prepared by preparations $\mathbb{P}_a$ and $\mathbb{P}_b$.  If measurements in  $\mathcal{M}$  are tomographically complete, then   
\be\label{kagg:App}
d_\text{trace}(\rho_a,\rho_b)\le  \mathbb{d}_\text{prep}(\mathbb{P}_a, \mathbb{P}_b) \ ,
\ee
where the  equality holds if preparations in $\mathcal{P}$ can prepare the  density operators 
\be \label{sigma:App}
 \sigma_{a/b}=\frac{\Pi_{a/b} (\rho_a-\rho_b) \Pi_{a/b}}{\Tr(\Pi_{a/b} (\rho_a-\rho_b))}\ ,
 \ee
where $\Pi_{a}$ and $\Pi_b$ are, respectively, projectors to the subspaces with non-negative and negative eigenvalues of  $\rho_a-\rho_b$.

To prove this statement, first we note that the assumption that measurements are tomographically-complete implies that any two equivalent preparations should be described by the same density operator.  Therefore, under this assumption, the definition in Eq.(\ref{second:App}) is equivalent to
 \begin{align}\label{efj}
\mathbb{d}_{\text{prep}}(\mathbb{P}_a,\mathbb{P}_b)\equiv \inf  
\Big\{r\ge 0: \exists \tau_{a},\tau_{b},  \rho_a-\rho_b= r [\tau_{b}-\tau_{a}]  \Big\}\ , 
\end{align}
where $\tau_{a}$ and $\tau_{b}$ are  density operators which can be prepared using preparations in $\mathcal{P}$. Taking the $l_1$ norm of both sides of  equation
\be
\rho_a-\rho_b= r [\tau_{b}-\tau_{a}]\ ,
\ee
and using the fact that 
\be
\|\tau_{b}-\tau_{a}\|_1\le \|\tau_{b}\|_1+\|\tau_{a}\|_1=2\ ,
\ee
 we find that $\| \rho_a-\rho_b\|_1\le 2r$, which implies  
\be\label{mbn}
\mathbb{d}_{\text{prep}}(\mathbb{P}_a,\mathbb{P}_b)\ge \frac{1}{2}\|\rho_a-\rho_b \|_1 = d_\text{trace}(\rho_a,\rho_b)\ . 
\ee
Next, assume preparations in $\mathcal{P}$ can prepare the pair of density operators 
\be
 \sigma_{a/b}=\frac{\Pi_{a/b} (\rho_a-\rho_b) \Pi_{a/b}}{\Tr(\Pi_{a/b} (\rho_a-\rho_b))}\ ,
 \ee
where $\Pi_{a}$ and $\Pi_b$ are, respectively, projectors to the subspaces with non-negative and negative eigenvalues of  $\rho_a-\rho_b$. Note that both $\sigma_{a}$ and $\sigma_{b}$ are positive operators, with trace one and hence valid density operators. Also, because $\Tr(\rho_a-\rho_b)=0$, then  
\be
\Tr(\Pi_a[\rho_a-\rho_b])=\Tr(\Pi_b[\rho_b-\rho_a])=\frac{1}{2}\|\rho_a-\rho_b \|_1\ .
\ee
Then, it can be easily seen that 
\be
\rho_a-\rho_b= \Tr(\Pi_a[\rho_a-\rho_b]) [\sigma_a-\sigma_b]= \frac{\|\rho_a-\rho_b \|_1}{2} [\sigma_{a}-\sigma_{b}]\ .
\ee
Comparing with the definition of the operational total variation distance in Eq.(\ref{efj}), we conclude that, if preparations in $\mathcal{P}$ can prepare both $\sigma_a$ and $\sigma_b$, then
\begin{align}
\mathbb{d}_{\text{prep}}(\mathbb{P}_a,\mathbb{P}_b)\le \frac{1}{2}\|\rho_a-\rho_b \|_1=d_\text{trace}(\rho_a,\rho_b)\  . 
\end{align}
Combining this with Eq.(\ref{mbn}), we find 
\begin{align}
\mathbb{d}_{\text{prep}}(\mathbb{P}_a,\mathbb{P}_b)= \frac{1}{2}\|\rho_a-\rho_b \|_1=d_\text{trace}(\rho_a,\rho_b)\  . 
\end{align}
 This completes the proof of proposition \ref{Thm0}.

 \subsection{Proof of Eq.(\ref{dso})}
In this section, we prove Eq.(\ref{dso}), i.e.,
\begin{align}\label{dso:App}
d_\text{TV}(\mu_a, \mu_b) &\le ({C}_\text{prep}+1)\times \mathbb{d}_\text{prep}(\mathbb{P}_a, \mathbb{P}_b)+ {C}_\text{prep}\ .
\end{align}

Suppose for $q\ge 0$, there exist preparations $\mathbb{P}_{a'}$ and $\mathbb{P}_{b'}$, such that
\be\label{wfwww}
\{(1-q, \mathbb{P}_a) , (q, \mathbb{P}_{a'}) \} \sim  \{(1-q, \mathbb{P}_{b}) , (q, \mathbb{P}_{b'}) \}\ .
\ee
Let  $\mu$ and $\mu'$  be, respectively, the probability distributions associated to the ensembles in the left-hand and the right-hand sides of this equality, i.e.,
\begin{align}
\mu &=(1-q) \mu_a+q  \mu_{a'}\\
\mu' &=(1-q) \mu_b+q  \mu_{b'}\ . 
\end{align}
Since the two ensembles are operationally indistinguishable, the total variation 
distance between their corresponding probability distributions is bounded by the inaccessible information of the model, i.e.
\begin{align}
{C}_\text{prep}&\ge d_\text{TV}(\mu,\mu')\\ 
&= d_\text{TV}\big((1-q)\mu_a+q \mu_{a'}, (1-q)\mu_b+q \mu_{b'}\big)\\ 
 &\ge (1-q) d_\text{TV}\big(\mu_a, \mu_b\big)- q\ d_\text{TV}\big(\mu_{a'}, \mu_{b'}\big)\\ &\ge (1-q)\ d_\text{TV}\big(\mu_a, \mu_b\big)- q  \ ,
 \end{align}
where the  third  line follows from the triangle inequality and the fourth line follows from the fact that the total variation distance is bounded by one. Dividing both sides by $1-q$, we find 
\begin{align}\label{lhwf902}
\frac{{C}_\text{prep}}{1-q}+\frac{q}{1-q}\ge d_\text{TV}(\mu_a,\mu_b)\ .
 \end{align}
Next, we take the infimum over all possible values of $q$ for which Eq.(\ref{wfwww}) holds for some of preparations $\mathbb{P}_{a'}$ and $\mathbb{P}_{b'}$. By definition the infimum of $q/(1-q)$ is  $\mathbb{d}_\text{prep}(\mathbb{P}_a, \mathbb{P}_b) $. Furthermore, since $\frac{1}{1-q}=1+\frac{q}{1-q}$, the infimum of   $\frac{1}{1-q}$ is $1+\mathbb{d}_\text{prep}(\mathbb{P}_a, \mathbb{P}_b)$. Therefore, Eq.(\ref{lhwf902}) implies
\begin{align}
{C}_\text{prep}\times (1+\mathbb{d}_\text{prep}(\mathbb{P}_a, \mathbb{P}_b))+\mathbb{d}_\text{prep}(\mathbb{P}_a, \mathbb{P}_b)\ge d_\text{TV}(\mu_a,\mu_b)\ ,
 \end{align}
or, equivalently, 
\begin{align}
({C}_\text{prep}+1)\times  \mathbb{d}_\text{prep}(\mathbb{P}_a, \mathbb{P}_b)+ {C}_\text{prep}\ge d_\text{TV}(\mu_a,\mu_b)\ ,
 \end{align}
which is Eq.(\ref{dso}).

\section{Inaccessible information for the Kochen-Spekker model of qubit}\label{App:E}

Recall that in Kochen-Specker model, for any pure state $|\psi\rangle,$ with Bloch vector $\hat{s}_\psi\in\mathbb{R}^3$, the corresponding probability density is
\be
\mu_\psi(\hat{n})=4  \hat{n}\cdot \hat{s}_\psi \times  \Theta(\hat{n}\cdot \hat{s}_\psi)\ ,
\ee
which satisfies the normalization
\be
\int \frac{d\Omega}{4\pi}\ \mu_\psi(\hat{n})=1\ ,
\ee
where $d\Omega$ is the solid angle differential. 
Here, $ \Theta$ is the Heaviside step function, and $\Theta(0)=1/2$. Similarly, for any two-outcome projective measurement with projectors $|\phi\rangle\langle\phi|$, and $I-|\phi\rangle\langle\phi|$, the response function associated to projector $\phi$ is
\be
\xi(\phi|\hat{n})= \Theta(\hat{n}\cdot \hat{r}_\phi)\ ,
\ee
where $\hat{r}_\phi$ is the Bloch vector corresponding to $\phi$.

Then, Kochen and Specker show that 
\begin{align}
\int \frac{d\Omega}{4\pi}\ \mu_\psi(\hat{n}) \xi(\phi|\hat{n})= \frac{1+ \hat{r}_\phi\cdot\hat{s}_\psi}{2}= |\langle\psi|\phi\rangle|^2 \ .
\end{align}
 Here, we find an upper bound on inaccessible information for this model.

Consider two ensembles of pure states $\{(p_i, \psi_i)\}$ and   $\{(p'_j, \psi'_j)\}$, which  are described by the same density operator, such that
\be
\sum_i p_i \psi_i=\sum_j p'_j \psi'_j\ .
\ee
This equation implies
\be
\sum_i p_i \hat{s}_i=\sum_j p'_j \hat{s}'_j \equiv \vec{a}\ ,
\ee
where $\hat{s}_i$ and $\hat{s}'_j$ are, respectively, the Bloch vectors of  $\psi_i$
and $\psi'_j$. The convex-linearity implies that  the probability distribution associated to the ensembles $\{(p_i, \psi_i)\}$ and  $\{(p'_j, \psi'_j)\}$ are, respectively, 
\begin{align}\label{lkje011}
\mu(\hat{n})&= \sum_i p_i \mu_i(\hat{n})=4\sum_i p_i {\hat{n}\cdot \hat{s}_i} \Theta(\hat{n}\cdot \hat{s}_i)=4\sum_i p_i \frac{\hat{n}\cdot \hat{s}_i+|\hat{n}\cdot \hat{s}_i|}{2}=2 (\hat{n}\cdot\vec{a}+\sum_i p_i |\hat{n}\cdot \hat{s}_i| )\ ,\\ \mu'(\hat{n})&= \sum_j p'_j \mu'_j(\hat{n})=4\sum_j p'_j {\hat{n}\cdot \hat{s}'_j} \Theta(\hat{n}\cdot \hat{s}'_j)=4\sum_j p'_j \frac{\hat{n}\cdot \hat{s}'_j+|\hat{n}\cdot \hat{s}'_j|}{2}=2 (\hat{n}\cdot\vec{a}+\sum_j p'_j |\hat{n}\cdot \hat{s}'_j| )\ \label{lkje012},
\end{align}
where we have used the fact that $x\times \Theta(x)=(x+|x|)/2$.  The total variation distance between $\mu$ and $\mu'$ can be written as
\begin{align}
d_\text{TV}(\mu,\mu')&\equiv\frac{1}{2}\int \frac{d\Omega}{4\pi}  \ |\mu(\hat{n})-\mu'(\hat{n})|\\ &=\frac{1}{2}\int \frac{d\Omega}{4\pi}\   \Big[\mu(\hat{n})+\mu'(\hat{n})-2\min\{\mu(\hat{n}), \mu'(\hat{n}) \}\Big] \\ &= 1- \int\frac{d\Omega}{4\pi}\  \min\{\mu(\hat{n}), \mu'(\hat{n}) \}\label{ewh}\ .
\end{align}
Therefore, using Eq.(\ref{lkje011}) and Eq.(\ref{lkje012}), we find
\bes\label{wgfg}
\begin{align}
d_\text{TV}(\mu,\mu')&=1-2\int \frac{d\Omega}{4\pi}\ \Big[{\hat{n}\cdot\vec{a}+\min\{\sum_i p_i |\hat{n}\cdot \hat{s}_i|, \sum_j p'_j |\hat{n}\cdot \hat{s}'_j| \}}\Big]\\ &=1-2 \int  \frac{d\Omega}{4\pi}\  \min\{\sum_i p_i |\hat{n}\cdot \hat{s}_i|, \sum_j p'_j |\hat{n}\cdot \hat{s}'_j| \}  \\ &=1-2\Big[\sum_i p_i \int_A \frac{d\Omega}{4\pi}  |\hat{n}\cdot \hat{s}_i|+\sum_j p'_j \int_{\overline{A}} \frac{d\Omega}{4\pi}  |\hat{n}\cdot \hat{s}'_j|\Big]\ ,
\end{align}
\ees
where to get the second line we have used $\int d\Omega\  \hat{n}=0$, and $A$ is the set of points on the unit sphere for which $\sum_i p_i |\hat{n}\cdot \hat{s}_i|\le \sum_j p'_j |\hat{n}\cdot \hat{s}'_j|$, i.e.,
\be
A=\{\hat{n}: \sum_i p_i |\hat{n}\cdot \hat{s}_i|\le  \sum_j p'_j |\hat{n}\cdot \hat{s}'_j|\ \}
\ee
and  $\overline{A}$ is its complement, i.e. the set of points for which 
$\sum_i p_i |\hat{n}\cdot \hat{s}_i|> \sum_j p'_j |\hat{n}\cdot \hat{s}'_j|$.

It can be easily shown that for any unit vector $\hat{s}$, and any subset $B$ of the unit sphere, 
\be\label{wgp2p}
\int_{B} \frac{d\Omega}{4\pi} \ |\hat{n}\cdot \hat{s}|\ge \frac{1}{2} \Big|\int_{B}\frac{d\Omega}{4\pi} \Big|^2\equiv\frac{\eta^2_B}{2}\ , 
\ee
where $\eta_B\equiv \int_{B}\frac{d\Omega}{4\pi}$ is the fraction of the total area of the unit sphere covered by region $B$ (To show this bound, we choose region $B$, such that $\int_{B} \frac{d\Omega}{4\pi} \ |\hat{n}\cdot \hat{s}|$ is minimized under the constraint  that the fraction of the total area covered by $B$ is $ \int_{B}\frac{d\Omega}{4\pi}=\eta_B$. Using the symmetry of the problem, the region $B$ for which this integral is minimized can be chosen to be symmetric under rotation around $\hat{s}$, and in the form $B=\{\hat{n}: |\hat{n}\cdot \hat{s}|< c\}$, where $c$ is chosen such that the constraint $\int_{|\hat{n}\cdot \hat{s}|< c} \frac{d\Omega}{4\pi}=\eta_B$ is satisfied. For this choice of region $B$ one can easily show that 
$\int_{B} \frac{d\Omega}{4\pi} \ |\hat{n}\cdot \hat{s}|= c^2/2$ and $ \int_{B}\frac{d\Omega}{4\pi}=c$. Setting $c=\eta_B$, we find $\int_{B} \frac{d\Omega}{4\pi} \ |\hat{n}\cdot \hat{s}|\ge \eta^2_B/2$).

Using the fact that $\overline{B}$ is the complement of $B$, and 
\be
\int_{B}\frac{d\Omega}{4\pi}+ \int_{\overline{B}}\frac{d\Omega}{4\pi}=1\ , 
\ee
we find that
\be
 \int_{\overline{B}}\frac{d\Omega}{4\pi}=1-\eta_B\ .
\ee
Therefore, applying Eq.(\ref{wgp2p}) for $\overline{B}$ instead of $B$, we find
\be
\int_{\overline{B}} \frac{d\Omega}{4\pi} \ |\hat{n}\cdot \hat{s}|\ge \frac{(1-\eta_B)^2}{2}\ .
\ee
Putting this into Eq.(\ref{wgfg}) and defining $\eta_A\equiv \int_{A}\frac{d\Omega}{4\pi}$, we find
\begin{align}
d_\text{TV}(\mu,\mu')&=1-2\Big[\sum_i p_i \int_A \frac{d\Omega}{4\pi}  |\hat{n}\cdot \hat{s}_i|+\sum_j p'_j \int_{\overline{A}} \frac{d\Omega}{4\pi}  |\hat{n}\cdot \hat{s}'_j|\Big]\ ,\\ 
&\le 1-2\Big[\sum_i p_i \frac{\eta^2_A}{2}+\sum_j p'_j \frac{(1-\eta_A)^2}{2} \Big]\\ 
&=1-\Big[{\eta^2_A} + (1-\eta_A)^2 \Big]\le \frac{1}{2}\ ,
\end{align}
where the last bound follows from the fact that $0\le \eta_A\le 1$, and $\eta^2_A + (1-\eta_A)^2$ is minimized for $\eta_A=1/2$.  This proves the inaccessible information for this model model is $C_\text{prep}\le \frac{1}{2}$. 

\section{Noisy Quantum Systems}

\subsection{Necessary and sufficient condition for Preparation Non-Contextuality\\ (Proof of the result in Sec.\ref{Sec:lkhdmain}) }\label{Sec:lkhd}

Consider the operational theory whose preparations prepare all and only states $\{\mathcal{E}(\rho)\}$ for arbitrary density operator $\rho$  of a quantum system,  and whose measurements $\mathcal{M}$ allow arbitrary measurements. Here, $\mathcal{E}$ is a completely positive trace-preserving map, which describes the noise process. 

Suppose the operational theory is described by a model satisfying   Preparation Non-Contextuality (PNC). PNC together with convex-linearity  imply that all preparations  which prepare a system in a density operator $\mathcal{E}(\rho)$, should be described by the same probability distribution, denoted by $\mu_{\mathcal{E}(\rho)}$. Therefore,  for any measurement $\mathbb{M}$ with POVM $\{B_l\}$, there exists a response function $\{\zeta_\mathbb{M}(l|\lambda): \zeta_\mathbb{M}(l|\lambda)\ge 0, \forall\lambda: \sum_l \zeta(l|\lambda)=1\}$, such that
\be\label{lkhqef}
\Tr(\mathcal{E}(\rho) B_l)=\sum_{\lambda}  \zeta_\mathbb{M}(l|\lambda)\times  \mu_{\mathcal{E}(\rho)}(\lambda)\ .
\ee
The fact that $\mu_{\mathcal{E}(\rho)}(\lambda)$  is a convex-linear function of density operator $\rho$,  implies that there exists a positive operator $E_\lambda$, such that $\mu_{\mathcal{E}(\rho)}(\lambda)=\Tr(E_\lambda \rho)$ (This follows from a generalization of Gleason's theorem in \cite{busch1999resurrection}, which has been  previously used in \cite{spekkens2005contextuality} to prove contextuality of quantum mechanics).  Furthermore, the normalization $\sum_\lambda \mu_{\mathcal{E}(\rho)}(\lambda)=1$, which holds for arbitrary density opertator $\rho$, implies $\sum_\lambda E_\lambda =I$.   We conclude that 
\be\label{lkhqef2}
\Tr(\mathcal{E}(\rho) B_l)=\sum_{\lambda\in \Lambda}   \zeta_\mathbb{M}(l|\lambda) \times \Tr(E_\lambda \rho)\  .
\ee
Next, we prove the converse. Suppose the probability of an outcome of any arbitrary measurement can be decomposed in the form of Eq.(\ref{lkhqef2}), for a fixed POVM $\{E_\lambda\}$. Then, based on this decomposition we can immediately define a model which satisfies both PNC and convex-linearity:  in this model, to any density operator $\mathcal{E}(\rho)$ we associate the  probability distribution $\mu_{\mathcal{E}(\rho)}(\lambda)\equiv \Tr(E_\lambda \rho)$. Combined with Eq.(\ref{lkhqef2}), this proves the claim. 

\subsection{Necessary and sufficient condition for both Preparation and Measurement Non-Contextualities \\ (Proof of the result in Sec.\ref{Sec:lkhd2main})  }\label{Sec:lkhd2}

Again, consider the  operational theory whose preparations prepare all and only states $\{\mathcal{E}(\rho)\}$ for arbitrary density operator $\rho$  of a quantum system  and whose measurements $\mathcal{M}$ allow arbitrary measurements. 

First, we show that if $\mathcal{E}$ is entanglement-breaking then this operational theory can be described by an ontological model satisfying  PNC, MNC and convex-linearity.

According to a well-known result by Holevo, any entanglement-breaking channel has a  decomposition as  
\be
\mathcal{E}(\tau)=\sum_{\gamma\in \Gamma} \Tr(F_\gamma \tau) \sigma_\gamma\ ,
\ee
 where  $\{F_\gamma\}_\gamma$ is a POVM  and $\{\sigma_\gamma\}_\gamma$ is a set of density operators.  Furthermore, assuming the Hilbert space is finite-dimensional, $\Gamma$ is a finite set.
 
 Based on this decomposition,  we can immediately define  an ontological  model for the operational theory. In this model  the ontic space is $\Gamma=\{\gamma\}$, and for each ontic state $\gamma\in\Gamma$, the probability associated to state $\mathcal{E}(\rho)$ is 
 \be
 \mu_{\mathcal{E}(\rho)}(\gamma)\equiv\Tr(F_\gamma \rho)\ .
 \ee
  Furthermore, for any  measurement $\mathbb{M}$ described by POVM  $\{B_l\}_l$, the response function associated to $B_l$ is 
  \be
  \zeta_\mathbb{M}(l|\gamma)\equiv\Tr(\sigma_\gamma B_l)\ .
  \ee
   Then, the probability of outcome $l$ for state $\mathcal{E}(\rho)$ can be written as $\Tr(\mathcal{E}(\rho)B_l)=\sum_{\gamma\in \Gamma}  \zeta_M(l|\gamma) \mu_{\mathcal{E}(\rho)}(\gamma)\ $, 
 which implies the model describes the operational theory. From the above definitions, we can easily see that this model satisfies convex-linearity, PNC and MNC.

Next, we prove the converse direction, i.e., we show that if there is a model  satisfying PNC, MNC and convex-linearity, then the noise channel $\mathcal{E}$ should be entanglement-breaking. As we saw in Sec.\ref{Sec:lkhd}, PNC and convex-linearity imply that there exists a POVM $\{E_\lambda\}$ such that for any measurement $\mathbb{M}$ with POVM $\{B_l\}_l$ holds that
\be
\Tr(\mathcal{E}(\rho) B_l)=\sum_{\lambda\in \Lambda}   \zeta_\mathbb{M}(l|\lambda) \times \Tr(E_\lambda \rho)\  ,
\ee
where $  \zeta_\mathbb{M}(l|\lambda)\ge 0$, and  $\forall \lambda:\ \sum_l  \zeta_\mathbb{M}(l|\lambda)=1$. 

Convex-linearity implies that   $ \zeta_\mathbb{M}(l|\lambda)$ is in the form $f_\lambda(B_l)$, where for each $\lambda$, $f_\lambda$ is a positive convex-linear function. Using the generalized Gleason's theorem \cite{busch1999resurrection} again, it follows that there exists a positive operator $\tau_\lambda$, such that  $\zeta_\mathbb{M}(l|\lambda)=\Tr(\tau_\lambda B_l) $ . Furthermore, the fact that $\sum_l \zeta_\mathbb{M}(l|\lambda)=1$ together with the fact that $\sum_l B_l=I$, implies that $\Tr(\tau_\lambda)=1$, i.e., $\tau_\lambda$ is a density operator. Therefore, $\Tr(\mathcal{E}(\rho)B_l)=  \sum_{\lambda\in \Lambda}  \Tr(\tau_\lambda B_l) \Tr(E_\lambda \rho)\ $ . Since this holds for an arbitrary state $\rho$ and positive operator $B_l$, we find $
\mathcal{E}(\cdot)= \sum_{\lambda\in \Lambda}  \tau_\lambda \Tr(\cdot B_\lambda)$, 
and therefore $\mathcal{E}$ is entanglement-breaking. This completes the proof of the statement in Sec.\ref{Sec:lkhd2main}. \\

\section{A preparation non-contextual ontological model for a noisy qubit } \label{Sec:ex-dep}

Consider the qubit deplorizing channel $\mathcal{D}_p$, defined by 
\be
\mathcal{D}_p(\rho)=(1-p) \rho+ p \frac{I}{2}\ ,
\ee
where $I/2$ is the maximally mixed state, and $0\le p\le 1$. Consider the operational theory whose preparations prepare all and only states $\{\mathcal{D}_p(\rho)\}$  for arbitrary density operator $\rho$  of a qubit  and whose measurements $\mathcal{M}$ allow arbitrary measurements. 
We show that for $p\ge 1/2$, this operational theory has a model satisfying PNC and convex-linearity.  In the following, first, we construct a model for the case of $p=1/2$. As we explain at the end, for $p>1/2$, we can construct a model by adding noise to this model. 

For $p=1/2$ this model is, in fact, a modified version of the Kochen-Specker model. Recall that  in the Kochen-Specker model  each ontic state is a point on the unit sphere,  which can be denoted by the unit vector $\hat{n}\in \mathbb{R}^3$. Then, for any pure state $\psi$, the corresponding probability density is 
\be
\mu_\psi(\hat{n})=4  \hat{n}\cdot \hat{s}_\psi \times  \Theta(\hat{n}\cdot \hat{s}_\psi)\ ,
\ee
where $\hat{s}_\psi$ is the Bloch vector associated to the density operator $\psi$. In the modified model, the probability density  associated to $\psi$ is
\be
\tilde{\mu}_\psi(\hat{n})=\hat{n}\cdot \hat{s}_\psi+1= 2|\langle\psi|\hat{n}\rangle|^2\ ,
\ee
which satisfies the normalization 
\be
\int\frac{d\Omega}{4\pi}\ \tilde{\mu}_\psi(\hat{n})=\int\frac{d\Omega}{4\pi} \ (\hat{n}\cdot \hat{s}_\psi+1)=1\ .
\ee
Note that $\tilde{\mu}_\psi(\hat{n})$ is in fact the probability density for outcome $\hat{n}$, when state $\psi$ is measured in a measurement described by the POVM $\{\frac{d\Omega}{2\pi} |\hat{n}\rangle\langle\hat{n}| \}$. 

The response function in this model is the same as the response function in the Kochen-Specker model. In particular, the response function associated to projector $\phi=|\phi\rangle\langle\phi|$ is 
\be
\xi(\phi|\hat{n})= \Theta(\hat{n}\cdot \hat{r}_\phi)\ ,
\ee
where $\hat{r}_\phi$ is the Bloch vector associated to $\phi$. In the following, we show that this model reproduces the statistics of the operational theory defined above, for $p=1/2$.  

Consider
\begin{align}\label{lllk}
\int \frac{d\Omega}{4\pi}\ \tilde{\mu}_\psi(\hat{n}) \xi(\phi|\hat{n})=\int \frac{d\Omega}{4\pi}\  \Theta(\hat{r}_\phi\cdot \hat{n})\  \times    2|\langle\psi|\hat{n}\rangle|^2=\langle\psi|M_\phi |\psi\rangle\ ,
\end{align}
where, operator $M_\phi$ is defined as 
\begin{align}
M_\phi=\int \frac{d\Omega}{4\pi}\   2 \Theta(\hat{n}\cdot \hat{r}_\phi)  |\hat{n}\rangle\langle\hat{n}|= f|\phi\rangle\langle\phi|+f' |\phi^\perp\rangle\langle\phi^\perp|\ ,
\end{align}
where $|\phi^\perp\rangle$ is the normalized vector orthogonal to $|\phi\rangle$, $f$ and $f'$ are non-negative real numbers and the equality follows from the symmetry of the integral.  
Using the normalization $\int \frac{d\Omega}{4\pi}=1$, we find 
\begin{align}
f+f'=\Tr(M_\phi)=\int \frac{d\Omega}{4\pi}\   2\ \Theta(\hat{n}\cdot \hat{r}_\phi) =1\ ,
\end{align}
which implies $f'=1-f$. We conclude that
\begin{align}
M_\phi= f|\phi\rangle\langle\phi|+(1-f) |\phi^\perp\rangle\langle\phi^\perp|=(2f-1)|\phi\rangle\langle\phi|+(1-f) I \ .
\end{align}
To calculate $f$, we consider Eq.(\ref{lllk}) for the special case $|\psi\rangle=|\phi\rangle$, which means $\hat{s}_\psi=\hat{r}_\phi$. 
Using the fact that  $\tilde{\mu}_\phi(\hat{n})=\hat{n}\cdot \hat{r}_\phi+1$, this equation implies
\begin{align}\label{lk2whf2}
f=\langle\phi|M_\phi |\phi\rangle=  \int \frac{d\Omega}{4\pi}\ \tilde{\mu}_\psi(\hat{n}) \xi(\phi|\hat{n})=\int \frac{d\Omega}{4\pi}\ [\hat{n}\cdot \hat{r}_\phi+1] \Theta(\hat{n}\cdot \hat{r}_\phi)=\frac{1}{2}+ \int \frac{d\Omega}{4\pi}\ (\hat{n}\cdot \hat{r}_\phi)  \Theta(\hat{n}\cdot \hat{r}_\phi)\ ,
\end{align}
where again we have used $\int \frac{d\Omega}{4\pi}=1$. 

To calculate the integral, we consider the case of $|\psi\rangle=|\phi\rangle$  in Kochen-Specker model, which implies 
\begin{align}
|\langle\phi|\phi\rangle|^2=\int \frac{d\Omega}{4\pi}\ \mu_\phi(\hat{n}) \xi(\phi|\hat{n})=\int \frac{d\Omega}{4\pi}\  4 (\hat{n}\cdot \hat{r}_\phi)  \Theta(\hat{n}\cdot \hat{r}_\phi) \ .
\end{align}
Therefore,
\be
\int \frac{d\Omega}{4\pi}\   (\hat{n}\cdot \hat{r}_\phi)  \Theta(\hat{n}\cdot \hat{r}_\phi) =\frac{1}{4}\ .
\ee
Together with Eq.(\ref{lk2whf2}), this implies $f=3/4$, i.e.,
\begin{align}
M_\phi= \frac{1}{2}|\phi\rangle\langle\phi|+\frac{1}{4} I \ .
\end{align}
Putting this into Eq.(\ref{lllk}), we find 
 \begin{align}
\int \frac{d\Omega}{4\pi}\ \tilde{\mu}_\psi(\hat{n}) \xi(\phi|\hat{n})=\Tr( |\psi\rangle\langle\psi|M_\phi)=\frac{1}{2} \Tr( |\psi\rangle\langle\psi||\phi\rangle\langle\phi|)+\frac{1}{4}=\Tr\Big(\mathcal{D}_{1/2}\big(|\psi\rangle\langle\psi|\big)|\phi\rangle\langle\phi|\Big)\ .
\end{align}
This proves that the ontological model correctly describes the statistics of
the operational theory whose preparations prepare all and only states in the form $\{\mathcal{D}_{1/2}(\rho)\}$, for arbitrary qubit density operator $\rho$. 

Since the probability distribution associated to density operator $|\psi\rangle\langle\psi|$, i.e., the density $\tilde{\mu}_\psi(\hat{n})=2\Tr(|\psi\rangle\langle\psi| |\hat{n}\rangle\langle\hat{n}|)$,  is a liner positive functional of the density operator  $|\psi\rangle\langle\psi| $,  this model is convex-linear and preparation non-contextual. 

Finally, note that for noise parameter $p>1/2$, we can obtain a non-contextual model by adding noise to this model. In particular, if instead of $\tilde{\mu}_\psi= \hat{n}\cdot \hat{s}_\psi+1$, we choose 
\be
\tilde{\mu}_\psi= 2(1-p) [\hat{n}\cdot \hat{s}_\psi+1] + (2p-1) \ ,
\ee  
then
 \begin{align}
\int \frac{d\Omega}{4\pi}\ \tilde{\mu}_\psi(\hat{n}) \xi(\phi|\hat{n}) &= 2(1-p)\Tr\Big(\mathcal{D}_{1/2}\big(|\psi\rangle\langle\psi|\big)|\phi\rangle\langle\phi|\Big)+\frac{2p-1}{2}\\ &= (1-p)\Tr\Big(|\psi\rangle\langle\psi| |\phi\rangle\langle\phi|\Big)+\frac{1-p}{2}+ \frac{2p-1}{2} \\ &= (1-p)\Tr\Big(|\psi\rangle\langle\psi| |\phi\rangle\langle\phi|\Big)+\frac{p}{2}\\ &=\Tr\Big(|\phi\rangle\langle\phi| \mathcal{D}_p(|\psi\rangle\langle\psi|) \Big) .
\end{align}

\end{document}